\documentclass[aps,twocolumn,amsmath,amssymb,floatfix,prx,reprint,footinbib,superscriptaddress,showpacs,longbibliography]{revtex4-1}

\usepackage{txfonts}
\usepackage{mathrsfs,times}
\usepackage{ulem}
\usepackage{textcomp}

\newcommand{\dg}{\dagger}
\newcommand{\nn}{\nonumber\\}

\usepackage{graphicx}
\usepackage{epstopdf}

\usepackage[applemac]{inputenc}
\usepackage[T1]{fontenc}
\usepackage[english]{babel}
\usepackage{ae}
\usepackage{siunitx}
\usepackage{color}
\usepackage{url}

\usepackage{amsmath,amssymb,natbib}
\usepackage{psfrag}
\usepackage{fixmath}
\usepackage{booktabs}

\usepackage{slashed}

\usepackage[americaninductors]{circuitikz}
\usepackage{tikz}
\usetikzlibrary{arrows}

\usepackage[colorlinks]{hyperref}
\hypersetup{%
        plainpages=true,
        breaklinks=true,
        hypertexnames=false,
        pageanchor=true,
        colorlinks=true,
        linkcolor={blue},
        citecolor={red},
        urlcolor={blue},
        anchorcolor={black}
      }

\newcommand{\be}{\begin{equation}}
\newcommand{\ee}{\end{equation}}
\newcommand{\bea}{\begin{eqnarray}}
\newcommand{\eea}{\end{eqnarray}}

\renewcommand{\eqref}[1]{\mbox{Eq.~(\ref{#1})}}

\begin{document}

\title{Floquet Many-body Engineering: Topological and Many-body Physics in Phase Space Lattices}

\author{Pengfei Liang}
\affiliation{Institut f\"ur Theoretische Festk\"orperphysik (TFP), Karlsruhe Institute of Technology (KIT), D-76131 Karlsruhe, Germany}
\affiliation{Department of Physics, Beijing Normal University (BNU), Beijing 100875, China}
\author{Michael Marthaler}
\affiliation{Institut f\"ur Theorie der Kondensierten Materie (TKM), Karlsruhe Institute of Technology (KIT), D-76131 Karlsruhe, Germany}
\affiliation{Theoretical Physics, Saarland University, 66123 Saarbr\"ucken, Germany}
\author{Lingzhen Guo}
\email{lzguo@tfp.uni-karlsruhe.de}
\affiliation{Institut f\"ur Theoretische Festk\"orperphysik (TFP), Karlsruhe Institute of Technology (KIT), D-76131 Karlsruhe, Germany}
\affiliation{Department of Microtechnology and Nanoscience (MC2), Chalmers University of Technology, SE-41296 G\"oteborg, Sweden}

\date{\today}

\begin{abstract}

Hamiltonians which are inaccessible in static systems can be engineered in periodically driven many-body systems, i.e., Floquet many-body systems.
We propose to use interacting particles in a one-dimensional (1D) harmonic potential with periodic kicking to investigate two-dimensional (2D) topological and many-body physics. Depending on the driving parameters, the Floquet Hamiltonian of single kicked harmonic oscillator has various lattice structures in phase space. The noncommutative geometry of phase space gives rise to the topology of the system. We investigate the effective interactions of particles in phase space and find that the point-like contact interaction in quasi-1D real space becomes a long-rang Coulomb-like interaction in phase space, while the hardcore interaction in pure-1D real space becomes a confinement quark-like potential in phase space. We also find that the Floquet exchange interaction does not disappear even in the classical limit, and can be viewed as an effective long-range spin-spin interaction induced by collision. Our proposal may provide platforms to explore new physics and exotic phases by \textit{Floquet many-body engineering}.
\end{abstract}


\maketitle

\section{Introduction}

Since the concept of topological order was first introduced into condensed matter physics in 1973 \cite{JPhysC1973}, topological phenomena have been intensively investigated in the past decades. Today, topology lies at the heart of many research fields, e.g., quantum Hall physics \cite{RevModPhys890250052017}, topological insulators/superconductors \cite{RevModPhys8230452010,RevModPhys8310572011}, and many more. The origin of topology in physics comes from the geometric phase factor of a quantum state when it moves along an enclosed path. In quantum Hall physics, the geometric phase is induced by the applied magnetic field and the resulting energy spectrum, also called the Hofstadter's butterfly \cite{PhysRevB1422391976}, is a fractal; while the band can be characterized by its topological invariant (Chern number or  TKNN invariant), which relates to the quantized Hall conductance directly \cite{PhysRevLett494051982}. In topological insulators/superconductors, the spin-orbit coupling takes the role of an effective magnetic field \cite{PhysRevLett952268012005,Science31417572006} resulting in the geometric quantum phase factor. For the ultracold atoms in optical lattice \cite{NaturePhysics1232005,RevModPhys808852008}, the geometric quantum phase (Berry phase \cite{Proc.R.Soc.Lond.A392451984}) is generated by shaking the lattice, which creates an artificial gauge field \cite{NewJournalofPhysics-5-56-2003,PRL-103-035301-2009,NJP-12-033041-2010,RevModPhys-83-1523-2011,RepProgPhys-77-126401-2014,Monika-Aidelsburger}.

\begin{figure}[!h]
  \centering
  \includegraphics[scale=0.3]{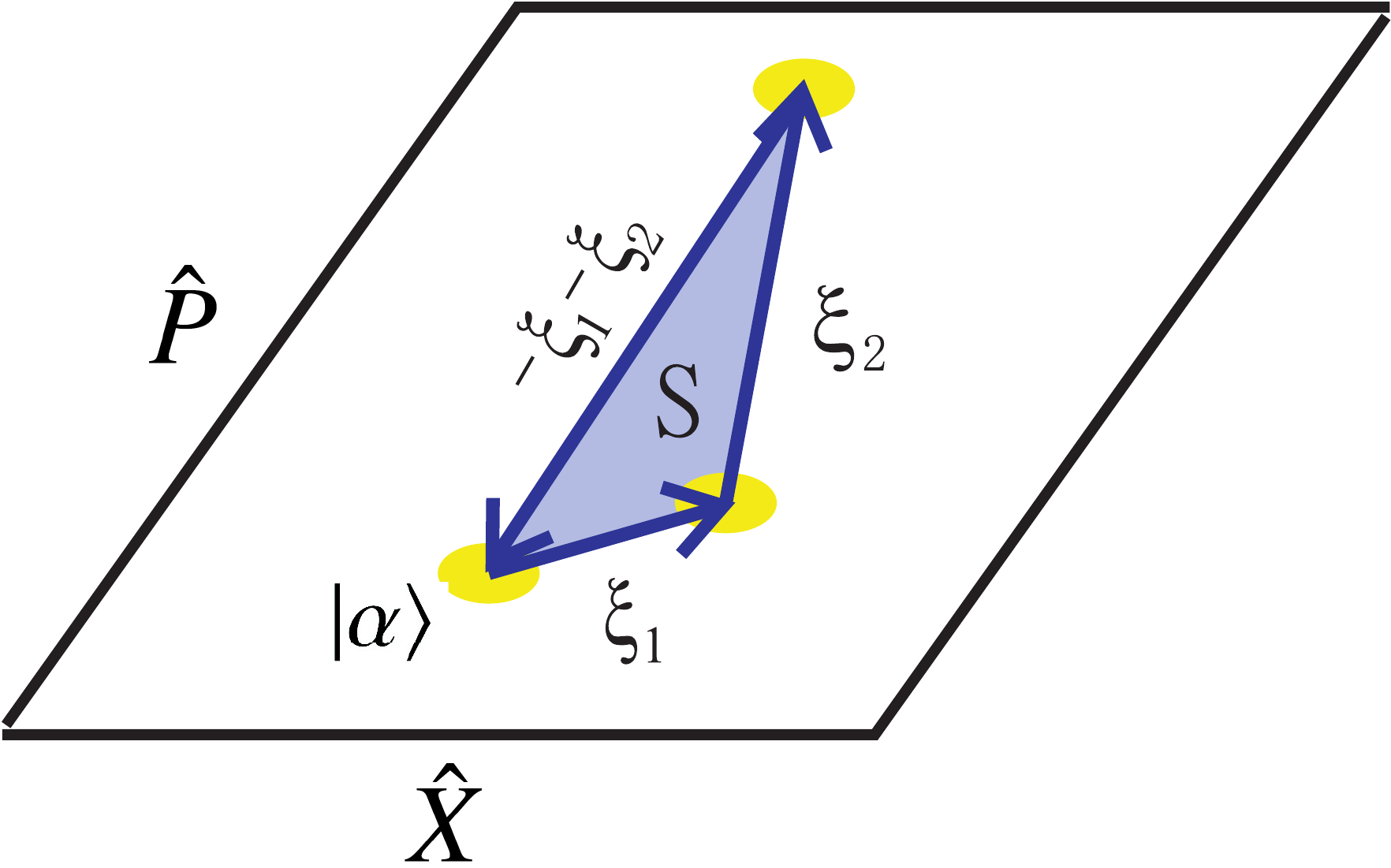}
 \caption{\label{fig_QunatumPhase}{\bf Geometric quantum phase in phase space.} A coherent state $|\alpha\rangle$ is moved along a closed triangle by three displacement operators, i.e., $D[-(\xi_1+\xi_2)]D(\xi_2)D(\xi_1)|\alpha\rangle=e^{i\frac{1}{\lambda}S}|\alpha \rangle,$ where $D(\xi)\equiv\exp(\frac{\xi}{\sqrt{2\lambda}} \hat{a}^\dagger-\frac{\xi^*}{\sqrt{2\lambda}}\hat{a})$ with two complex numbers $\xi_1, \ \xi_2$ determining the moving path. The geometric phase factor is given by $e^{i\frac{1}{\lambda}S}$ with $S=\frac{1}{2}\mathrm{Im}[\xi_2\xi^*_1]$ being the area of the enclosed path (blue area). }\label{}
\end{figure}

An alternative way to study topological physics is employing the noncommutativity of phase space in quantum mechanics. In a noncommutative space, the concept of point is meaningless due to the commutative relationship $[\hat{X},\hat{P}]=i\lambda$. Instead, we should define a coherent state $|\alpha\rangle$ which is the eigenstate of the lowering operater, i.e., $\hat{a}|\alpha\rangle=\alpha|\alpha\rangle$ with $\hat{a}\equiv(\hat{X}+i\hat{P})/\sqrt{2\lambda}$. As shown in Fig.~\ref{fig_QunatumPhase}, we observe that a coherent state moving along a closed path in phase space naturally acquires an additional quantum phase factor  $e^{iS/\lambda}$, where $S$ is the enclosed area \cite{PRL1081704012012}. This observation reveals the origin of topology in the study of many dynamical systems, e.g., the kicked harmonic oscillator (KHO) \cite{MZas2005,ZhEkspTeorFiz-91-500-1986,Nonlinearity-4-543-1991,PRL932041012004} and the kicked Harper model (KHM) \cite{PhysRevLett-69-3302-1992,IntJModPhysB-08-207-1994}. The energy spectra of these dynamical systems exhibit butterfly structure and band topology similar to quantum Hall systems \cite{PhysRevA-80-023414-2009,PhysRevLett-67-3635-1991}. In the strong kicking strength regime, the dynamical systems become chaotic and exhibits many novel behaviors such as dynamical localization, which has an intimate relation with the topology of bands \cite{PhysRevLett-65-3076-1990,Chaos-2-125-1992,CanadianJournalofChemistry-92-77-2014}.

In many-body physics of equilibrium systems, many exotic phases of matter emerge when interaction makes the system strongly correlated. It is the interplay between topology and interaction that gives rise to the fractional quantum Hall effect \cite{PhysRevLett48-1559-1982,PhysRevLett-50-1395-1983,RevModPhys-71-875-1999}, and many other fascinating phenomena \cite{PhysRevB-43-8641-1991,PhysRevLett-106-236802-2011,PhysRevLett-106-236804-2011,PhysRevLett-108-126405-2012}, like  fractional charge and anyons \cite{ILNuovoCimentoB-37-1-23-1977,PhysRevLett-49-957-1982,PhysRevLett-66-802-1991,PhysRevB-72-075342-2005,Nature-464-187-2010,Avinash-Khare}. Alternatively, it is also possible to engineer novel phases in periodically driven systems, i.e., the \textit{Floquet systems}. The Hamiltonian of a Floquet system is a periodic function in time, i.e., $H(t)=H(t+T)$. The Floquet theory \cite{PhysRev-138-B979-1965,PhysicsReports3042291998} allows us to describe stroboscopic time-evolution for every period by a time-independent Hamiltonian which is called the \textit{Floquet Hamiltonian} $H_F$ and is defined by $e^{-\frac{i}{\hbar}TH_F}\equiv\mathcal{T}e^{-\frac{i}{\hbar}\int_0^TH(t)dt}$, or equivalently
\begin{equation}\label{FloquetH}
H_F=i\frac{\hbar}{T}\mathrm{ln}\Big[\mathcal{T}e^{-\frac{i}{\hbar}\int_0^TH(t)dt}\Big].
\end{equation}
Here, $T$ is the chosen stroboscopic time step and $\mathcal{T}$ is the time-ordering operator. Exotic Floquet Hamiltonians \cite{PRL-91-110404-2003,PRA-68-013820-2003,PRL-111-175301-2013,PRE-91-032923-2015,NJP-093039-2015,PRL-115-075301-2015,PRA-93-053616-2016} which are inaccessible in static systems can be engineered from Eq.~(\ref{FloquetH}) and a range of novel physical phenomena, such as phase space crystals \cite{PRL-111-205303-2013,NewJPhys-18-0230065-2016}, Anderson localization (or many-body localization) in time domain \cite{SciRep-5-10787-2015,PRA-94-023633-2016,PRA-95-063402-2017,PRB-96-140201-2017} and spontaneous breaking of discrete time-translation symmetry (Floquet time crystals) \cite{PRE-82-031134-2010,PRA-91-033617-2015,PRL-117-090402-2016,PRL-116-250401-2016,PRL-118-030401-2017,Nature-543-217-2017,Nature5432212017,arXiv170403735,arXiv170207931}, can be created by Floquet engineering \cite{AIP-64-139-2015,JPBAMOP-49-013001-2016,RMP890110042017}. While most work focus on the single-particle physics of (dissipative) Floquet systems, the possible new physics by \textit{Floquet many-body engineering} has become an active research direction in recent years. Unlike the static many-body systems, the generic nonintegrable Floquet many-body systems are expected to be heated up, by the driving field, to a trivial stationary state with infinite temperature \cite{PRX-4-041048-2014,PRE-90-012110-2014,AOP3531962015}. However, before reaching the long-time featureless infinite-temperature state, there is a prethermal state with exponentially long lifetime for high driving frequencies, and therefore existing a prethermal dynamics which can be described by the time-independent Floquet Hamiltonian (\ref{FloquetH}) \cite{PRL1152568032015,PRL1161204012016,CMP3548092017,AOP367962016,PRB931551322016,PRE930121302016,PRB950141122017,SciRep7453822017,arXiv1706.07207,arXiv1708.01620}. By introducing disorder as in many-body localized systems \cite{NaturePhysics134602017} or coupling the Floquet many-body system to a cold bath \cite{PRX70110262017}, it is also possible to protect the metastable prethermal state.

In this paper, we investigate  cold atoms trapped in 1D harmonic potential with a stroboscopically applied optical lattice. The equation of motion of a single atom corresponds
to the kicked harmonic oscillator (KHO) and we find that
that intriguing 2D topological and many-body physics emerges in phase space.
The Floquet Hamiltonian of a single KHO, in the rotating wave approximation (RWA), forms various lattice structures in phase space depending on the driving parameters. The full dissipative quantum dynamics shows that the stationary state forms a lattice structure in phase space but with a finite size limited by the dissipation rate. Furthermore, we consider the interaction between cold atoms and find that the point-like contact interaction of cold atoms in real space becomes a long-range Coulomb-like interaction in phase space. More interestingly, the hard-core interaction of cold atoms in real space becomes a long-range potential which increases linearly with the distance in phase space, i.e., a quark-like confinement potential. We also find the Floquet exchange interaction has Coulomb-like long-range behavior, which does not disappear in the classical limit and becomes an effective spin-spin interaction.

\section{Model and Hamiltonian}

We start from interacting cold atoms trapped by an elongated three-dimensional harmonic potential, with the radial motion cooled down to the ground state. In this way, the spatial motion of the atoms is restricted to the remaining axial direction. In general, the one-dimensional system is described by
\begin{equation}\label{ManybodyH}
H(t)=\sum_{i}H_s(\hat x_i,\hat p_i,t)+\sum_{i<j}V(\hat x_i-\hat x_j),
\end{equation}
where $V(\hat x_i-\hat x_j)$ is the two-body interaction, which is typically contact or hard-core interactions in the context of cold atoms \cite{RevModPhys808852008,PRL819381998,PRL911632012003,PRL920304022004,Nature4292772004,Science30511252004,Science32512242009}. The $H_s(\hat{x}_i,\hat{p}_i,t)$ is the single-particle Hamiltonian which can be explicitly time-dependent. Here, the single-particle Hamiltonian is the quantum kicked harmonic oscillator, which is described by
$
H_s=\frac{1}{2m}\hat{p}_i^2+\frac{1}{2}m\omega_0^2\hat{x}_i^2+K_0T_d\cos (k\hat x_i)\sum_{n=-\infty}^{\infty}\,\delta(t-nT_d),
$
where $\omega_0$ is the axial harmonic frequency and $m$ is the atom's mass. The periodic term is implemented by a stroboscopic optical lattice, which can be created by two counter-propagating laser beams with off-resonant frequency far away from internal electronic transitions \cite{NaturePhysics1232005,RevModPhys808852008}. Parameters $k$ and $K_0$ are the wave vector of the laser beams and the kicking strength, respectively. Parameter $T_d$ is the time period between adjacent kicking pluses. We scale the coordinate and momentum by the units of $\sqrt{\hbar/(\lambda m\omega_0)}$ and $\sqrt{m\hbar\omega_0/\lambda}$ with the parameter $\lambda\equiv\hbar k^2/m\omega_0$, respectively. Finally, we have the dimensionless single-particle Hamiltonian scaled by $\hbar\omega_0/\lambda$
\begin{equation}\label{eq:dimensionlessKHO}
H_s(\hat x_i,\hat p_i,t)=\frac12(\hat{x}_i^2+\hat{p}_i^2)+K\tau\cos\hat{x}_i\sum_{n=-\infty}^{\infty}\,\delta(t-n\tau),
\end{equation}
where $K=\lambda K_0/\hbar\omega_0$ is the dimensionless kicking strength, $\tau=\omega_0 T_d $ is the dimensionless kicking period and the time $t$ has also been scaled by $\omega_0^{-1}$. The commutation relationship between the coordinate and the momentum is now $[\hat{x}_i,\hat{p}_j]=i\lambda\delta_{ij}$, where the dimensionless parameter $\lambda$ plays the role of an effective Planck constant. Thus, the semiclassical regime corresponds to the limit $\lambda\rightarrow0$. Accordingly, the two-body interaction will be given by the new scaled dimensionless observables as $V(\hat x_i-\hat x_j)$.

Our remaining paper is organized as follows. In Sec.~\ref{sec:PSCrystals}, we discuss the single-particle physics neglecting interaction of paticles. We first introduce the topological band theory of phase space lattices in Sec.~\ref{sec:BandDiagram}. Then, in Sec~\ref{sec:Dissipative}, we investigate the dissipative quantum dynamics of a KHO in a realistic environment and show how a lattice structure is formed in phase space. In Sec.~\ref{sec:Manybody}, we consider the interactions and investigate the many-body dynamics. We first develop a general theory of transforming a given real space interaction potential to a phase space interaction potential in Sec~\ref{sec:PhaseSpaceInteraction}. Then, in Sec~\ref{sec:Applications}, we apply our theory of phase space interaction to the special cases of contact interaction and hard-core interaction of cold atoms, and give the analytical expressions of corresponding phase space interactions. In Sec~\ref{Sec:classicalmanybody} and \ref{sec:DynamicalCrystals}, we investigate the many-body dynamics in the classical limit and discuss the concept of dynamical crystals. Finally, we summarize our results in Sec.~\ref{Sec:SummaryandOutlook}.

\begin{figure}
\centering
\includegraphics[width=\linewidth]{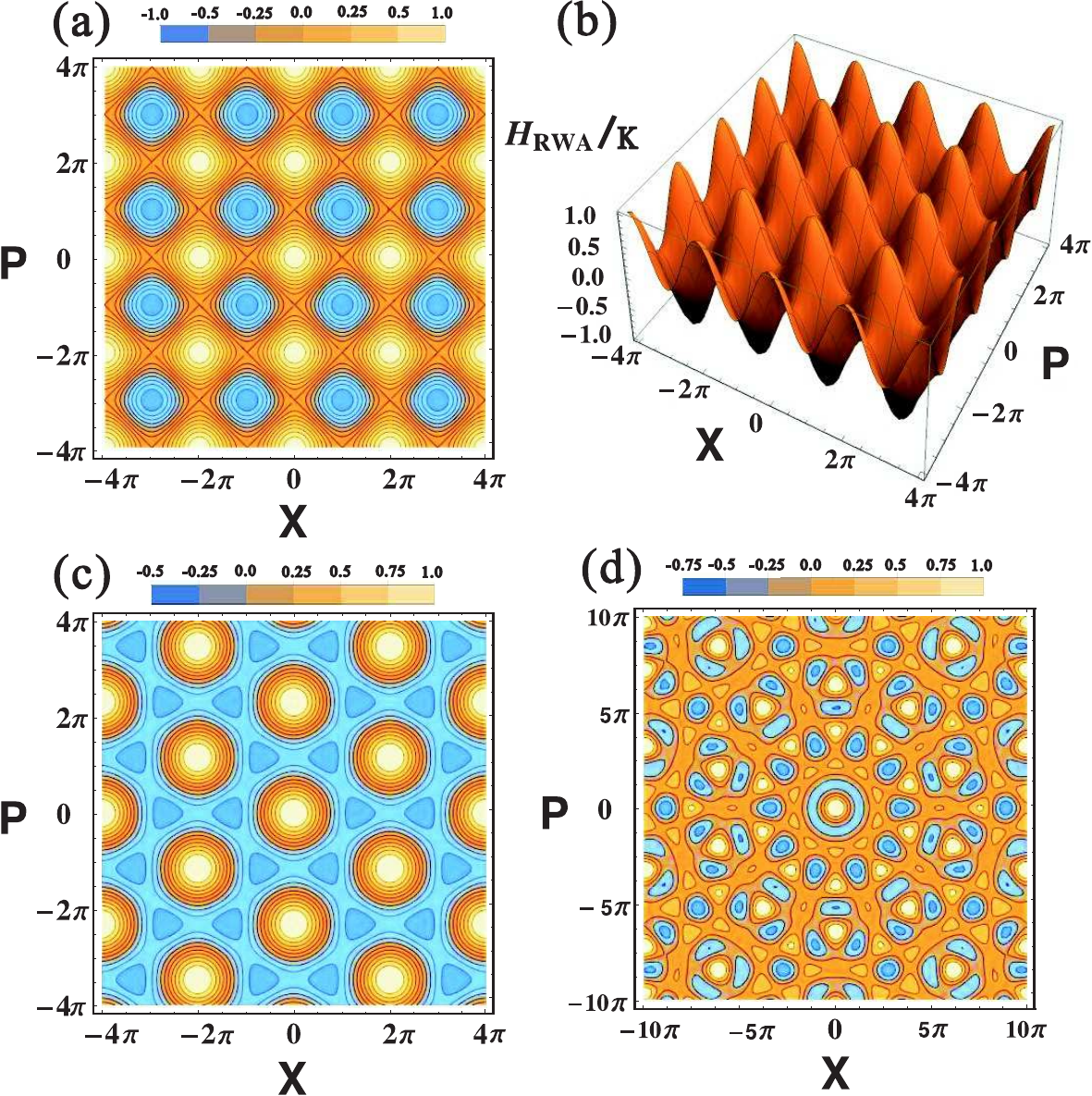}
\caption{\label{fig:psl}{\bf Phase space lattices:} $H_{RWA}(X,P)$ for different $q_0$.
(a) 2D density plot of square lattice for $q_0=4$. (b) 3D plot of square lattice for $q_0=4$. (c) Hexagonal lattice for $q_0=3$ or $q_0=6$. (d) Quasicrystal structures for $q_0=5$. The value of $H_{RWA}(X,P)$ has been scaled by the kicking strength $K$ in all figures.}
\end{figure}

\section{Phase Space Lattices}\label{sec:PSCrystals}

In this section, we investigate the single-particle Hamiltonian of the quantum KHO, i.e., \eqref{eq:dimensionlessKHO}, in the resonant condition that the kicking period satisfies $\tau=2\pi/q_0$ with $q_0$ an integer. When the kicking strength is weak $|K|\ll 1$, the single-particle dynamics is still dominated by the fast harmonic oscillation. Then we transform into an appropriately chosen rotating frame generated by the free time-evolution operator $\hat{O}(t)\equiv \exp(i\sum_{i}\hat{a}_i^\dagger\hat{a}_it/\lambda)$, where $\hat{a}_i$ is the annihilation operator defined by $\hat{a}_i\equiv(\hat{x}_i+i\hat{p}_i)/\sqrt{2\lambda}$. We transform the coordinates and momenta of particles by
\begin{eqnarray}\label{xpXP}
\left\{
\begin{array}{lll}
\hat{O}(t)\hat{x}_i\hat{O}^\dagger(t)&=&\hat{P}_i\sin t+\hat{X}_i\cos t\\
\hat{O}(t)\hat{p}_i\hat{O}^\dagger(t)&=&\hat{P}_i\cos t-\hat{X}_i\sin t.\\
\end{array}
\right.
\end{eqnarray}
Here, the operators $\hat{X}_i$ and $\hat{P}_i$ describe the dynamics of the $i$-th atom's phase and amplitude. For the harmonic oscillator, $\hat{X}_i$ and $\hat{P}_i$ are fixed and correspond to the initial state of $\hat{x}_i(t)$ and $\hat{p}_i(t)$. In our case, however, the phase and amplitude of KHO are slightly changed every harmonic time period due to the weak kicking. The time-evolution of $\hat{X}_i$ and $\hat{P}_i$ is slow compared to the fast global harmonic oscillation and can be obtained stroboscopically from the time-evolution of $\hat{x}_i(t)$ and $\hat{p}_i(t)$ every time period of $2\pi$.

From Eq.~(\ref{xpXP}), we have $[\hat{X}_i,\hat{P}_j]=[\hat{x}_i,\hat{p}_j]=i\lambda\delta_{ij}$. The canonical transformation of the single-particle Hamiltonian is given by $\hat{O}(t)H_s\hat{O}^\dagger(t)-iO(t)\dot{O}^\dagger(t)$. In RWA, we drop the fast oscillating
terms and arrive at the time-independent Hamiltonian (see detailed derivation in Appendix \ref{app:HRWA})
\begin{equation}\label{eq:RWAHamiltonian}
H_{\it RWA}(\hat X,\hat P)=\frac{K}{q_0}\sum_{j=1}^{q_0}\,\cos\left(\hat X\cos\frac{2\pi j}{q_0}+\hat P\sin\frac{2\pi j}{q_0}\right).
\end{equation}
Here, we have dropped the index of the operators since we are considering single-particle physics. Another way of deriving $H_{\it RWA}(\hat X,\hat P)$ is based on the series expansion of the Floquet Hamiltonian (\ref{FloquetH}) in order of the kicking strength $K$. By replacing the Planck constant $\hbar$ by a dimensionless one $\lambda$ and choosing the stroboscopic time step $T=q_0\tau$, we have the time-evolution operator in one stroboscopic time step
$\mathcal{T}e^{-\frac{i}{\lambda}\int_0^{q_0\tau}H_s(t)dt}=\hat{F}^{q_0}$ with the Floquet operator for one period $\hat F\equiv e^{-i(\hat{x}^2+\hat{p}^2)\tau/2\lambda}e^{iK\cos{\hat{x}}/\lambda}$. In the Appendix \ref{app:HRWA}, we show that $H_{\it RWA}(\hat X,\hat P)$ is indeed the first order expansion of the Floquet operator $\hat{F}^{q_0}$.

To display the symmetries of $H_{\it RWA}(\hat X,\hat P)$ in phase space, we calculate the averaged $H_{\it RWA}(\hat X,\hat P)$ in the coherent state representation (see the details in Appendix~\ref{app:Hdiag}), i.e., $$\langle \alpha|H_{\it RWA}(\hat X,\hat P)|\alpha\rangle=e^{-\lambda/4}H_{\it RWA}(X,P).$$ Here, the coherent state $|\alpha\rangle$ is defined by the eigenstate of the annihilation operator $\hat{a}\equiv(\hat{X}+i\hat{P})/\sqrt{2\lambda}$, i.e., $\hat{a}|\alpha\rangle=\alpha|\alpha\rangle$. The averaged position and momentum are $X\equiv\langle\alpha|\hat X|\alpha\rangle=\sqrt{2\lambda}\mathrm{Re}[\alpha]$ and $P\equiv\langle\alpha|\hat P|\alpha\rangle=\sqrt{2\lambda}\mathrm{Im}[\alpha]$. The quantity $H_{\it RWA}(X,P)$ has the same expression as \eqref{eq:RWAHamiltonian} but replacing the operators $\hat X$ and $\hat P$ by the averaged values $X$ and $P$ respectively. In Fig.~\ref{fig:psl}, we plot $H_{\it RWA}(X,P)$ in phase space for different $q_0$. We see that the $H_{\it RWA}(X,P)$ has a square lattice structure for $q_0=4$, hexagonal lattice structure for $q_0=3$ or $q_0=6$, and even quasicrystal structure for $q_0=5$ or $q_0\geq7$. The translational symmetry of the Hamiltonian~(\ref{eq:RWAHamiltonian}) in phase space gives rise to the band structure of its spectrum.


%
\begin{figure*}
\centering
\includegraphics[width=\linewidth]{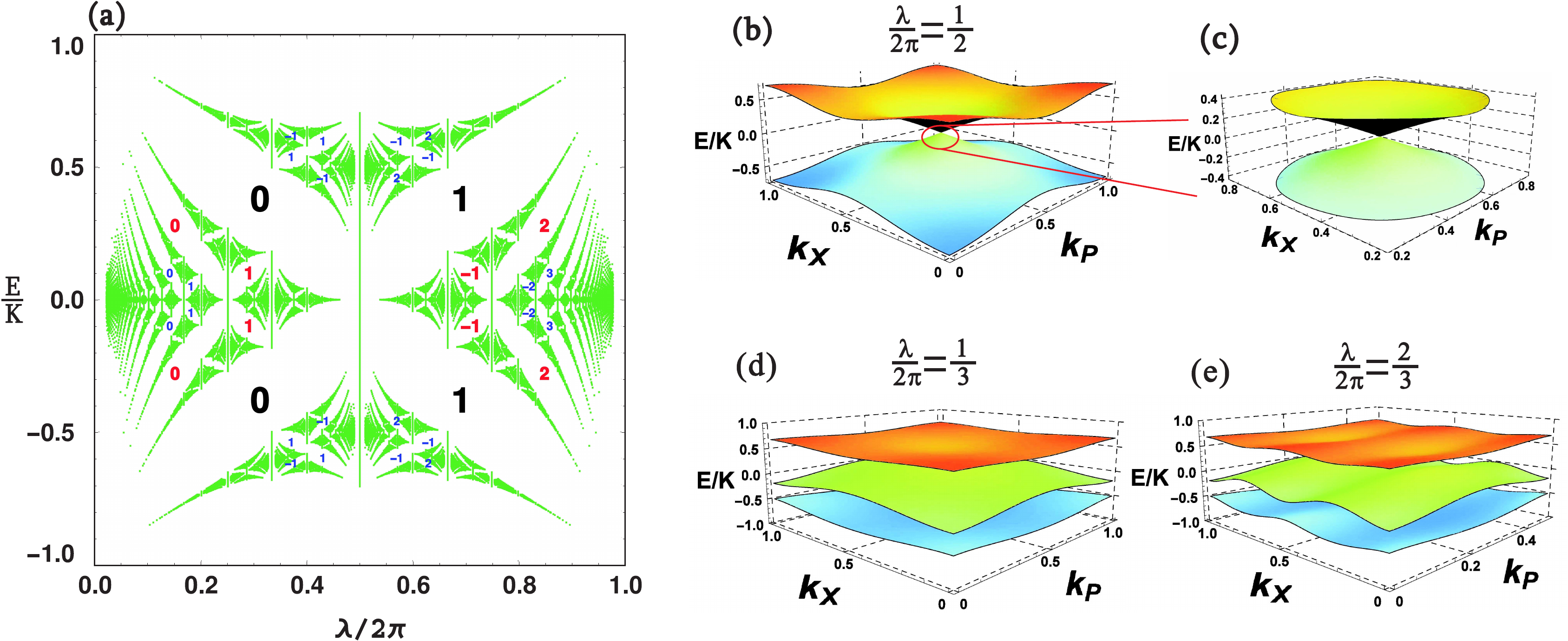}
\caption{\label{fig:bands}{\bf Quasienergy band structures.} (a) Hofstadter's butterfly: the quasienergy spectrum of Hamiltonian (\ref{eq:Hamiltoniansq}) with rational number $\lambda/2\pi\in[0,1]$. (b)-(e) Quasienergy band structures in 2D Brillouin zone for different parameters $\lambda$ which are given above the figures. For $\lambda/2\pi=1/2$, the two bands touch each other in the center of Brillouin zone. The linear dispersion relationship near the touching point is shown in figure (c).  }
\end{figure*}

\subsection{Band Structure and Topology}\label{sec:BandDiagram}
We will deal with the case of square lattice ($q_0=4$) in detail but the results can be readily generalized to the case of hexagonal lattice ($q_0=3$ or $6$). For $q_0=4$, the effective Hamiltonian (\ref{eq:RWAHamiltonian}) is further simplified as
\begin{eqnarray}\label{eq:Hamiltoniansq}
H_{sq}(\hat X,\hat P)=\frac 12K\Big(\cos\hat X+\cos\hat P\Big).
\end{eqnarray}
This Hamiltonian is closely related to the established Harper's equation, which is a tight binding model governing the motion of noninteracting electrons in the presence of a two-dimensional periodic potential and a uniformly threading magnetic field \cite{Harper1955,PhysRevB1422391976}.
The $H_{sq}(\hat X,\hat P)$ is invariant under discrete translation in phase space by two operators
$\hat{T}_1\equiv e^{i\frac{2\pi}{\lambda}\hat{P}}$ and $\hat{T}_2\equiv e^{i\frac{2\pi}{\lambda}\hat{X}}$, i.e.,
\begin{eqnarray}
\left\{
\begin{array}{lll}
\hat{T}_1H_{sq}(\hat X,\hat P)\hat{T}^\dagger_1=H_{sq}(\hat X+2\pi,\hat P)=H_{sq}(\hat X,\hat P)\\
\hat{T}_2H_{sq}(\hat X,\hat P)\hat{T}^\dagger_2=H_{sq}(\hat X,\hat P+2\pi)=H_{sq}(\hat X,\hat P).\\
\end{array}
\right.
\end{eqnarray}
The translation operators $\hat{T}_1$ and $\hat{T}_2$ generate an invariance group $G$ of $H_{sq}$ \cite{Dana1995}, which is a nonabelian group due to the identity $[\hat{T}^r_1,\hat{T}^s_2]=\hat{T}_1\hat{T}_2(1-e^{-i4\pi^2rs/\lambda})$ with integer powers $r,s\in\mathbb{Z}$. However, the group $G$ has abelian subgroups $G_a$ generated by $\hat{T}_1^r$ and $\hat{T}_2^s$ if $2\pi rs/\lambda\in \mathbb{Z}$, which means the value of the parameter $\lambda/2\pi$ needs to be a rational number, i.e., $\lambda/2\pi=p/q$, where $p$ and $q$ are coprime integers. Here, we choose the abelian subgroup $G_a$ generated by the following two generators ($r=1$, $s=p$)
\begin{equation}\label{eq:subgroup}
\hat{T}_X\equiv\hat{T}_1=e^{i\frac{2\pi }{\lambda}\hat{P}},    \ \ \ \ \ \     \hat{T}_P\equiv\hat{T}^p_2=e^{i\frac{2\pi p}{\lambda}\hat{X}}.
\end{equation}
Therefore, we can find the common eigenstates of commutative operators $\hat{T}_X$ and $\hat{T}_P$ with eigenvalues given by $e^{i2\pi k_X}$ and $e^{i2\pi p k_P}$ respectively. The boundaries of the two dimensional Brillouin zone are defined by $0\leq k_X\leq1$ and $0\leq k_P\leq1/p$,
where $k_X$ and $k_P$ are quasimomentum and quasicoordinate, respectively \cite{Zak1972}. The corresponding eigenvalues of the Hamiltonian $H_{sq}$ are also called \textit{quasienergies}.


%
%

The discrete translational symmetry in phase space allows us to determine the quasienergy spectrum numerically in Zak's $kq$-representation (see the instruction in Appendix~\ref{sec:appendixC} or Ref.~\cite{Zak1972}). Given the parameters of $\lambda/2\pi=p/q$, $k_X$ and $k_P$, the eigenvalues $E$ of $H_{sq}$ are determined by the following polynomial equation (see the derivation in Appendix~\ref{app:butterfly} or Ref.~\cite{Brown1968})
\begin{equation}\label{eq:butterfly}
\cos(q\lambda k_X)+\cos(q\lambda k_P)=1+\frac{1}{2}\mathrm{Tr}\prod_{j=1}^q
\begin{bmatrix}
    \frac{4E}{K}-2\cos(j\lambda) & -1 \\
    1 & 0
\end{bmatrix}.
\end{equation}
The left hand side of Eq.~(\ref{eq:butterfly}) takes values in the range $[-2,2]$ when the quasimomentum $k_X$ and quasicoordinator $k_P$ run over the whole Brillouin zone. The right hand side of Eq.~(\ref{eq:butterfly}) is a periodic function of $\lambda$ with period $2\pi$. Therefore, the quasienergy spectrum is also a periodic function of $\lambda$ with period $2\pi$. In Fig.~\ref{fig:bands}(a), we plot the quaienergy spectrum for $\lambda/2\pi\in[0,1]$, showing a Hofstadter butterfly structure identical to that in quantum Hall systems. In Fig.~\ref{fig:bands}(b), (d) and (e), we plot the quasienergy band structures in the two-dimensional Brillouin zone $(k_X,k_p)$ for $\lambda/2\pi=1/2, 1/3$ and $2/3$, respectively. For the given parameter $\lambda/2\pi=1/2$, we can obtain the analytical solutions from \eqref{eq:butterfly}, i.e., $$E=\pm \frac{1}{2}K\sqrt{1+\frac{1}{2}(\cos2\pi k_X+\cos2\pi k_P)}.$$ The two bands touch each other at the central point of the Brillouin zone, i.e., $(k_X=\frac{1}{2},k_P=\frac{1}{2})$, where the dispersion relationship becomes linear near the touching point, i.e., $E\approx\pm\frac{\pi K}{\sqrt{2}}|k|$ with $|k|\equiv\sqrt{(k_X-\frac{1}{2})^2+(k_P-\frac{1}{2})^2}$, as shown in Fig.~\ref{fig:bands}(c). In general, the two innermost bands always touch each other for even integer $q$. We also see that the quasienergy band structure is two-fold degenerate for $\lambda/2\pi=2/3$ while there is no degeneracy for $\lambda/2\pi=1/2$ and $\lambda/2\pi=1/3$. In fact, for each rational $\lambda/2\pi=p/q$ (remembering $p,q$ are coprime integers), the spectrum contains $q$ bands and each band has a $p$-fold degeneracy due to the fact that the invariance group $G$ can be expressed as the coset sum $\sum_{r=1}^p\hat{T}_1^rG_a$ \cite{Dana1995}.

\begin{figure}
\centering
\includegraphics[width=\linewidth]{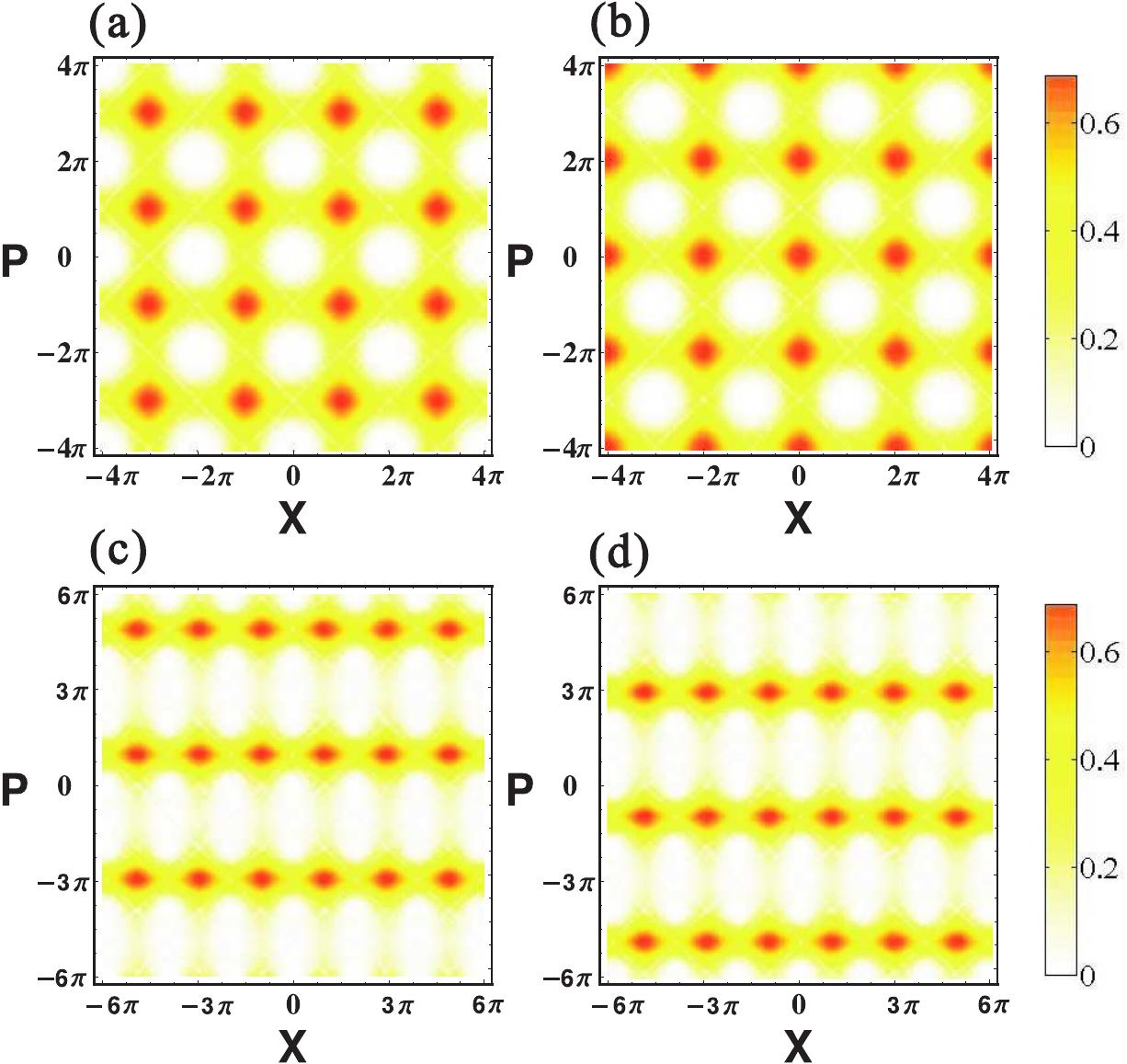}
\caption{\label{fig:Qfunctions}{\bf Husimi $Q$-functions of eigenstate $|\psi_{b,{\mathbf k}}\rangle$.} (a) Eigenstate $|\psi_{1,(0,0)}\rangle$ with the minimum quasienergy of the lower band for $\frac{\lambda}{2\pi}=\frac{1}{2}$. (b) Eigenstate $|\psi_{2,(0,0)}\rangle$ with the maximum quasienergy of the upper band for $\frac{\lambda}{2\pi}=\frac{1}{2}$. (c) Eigenstate $|\psi_{1,(\frac{1}{4},\frac{1}{2})}\rangle$ for $\frac{\lambda}{2\pi}=\frac{2}{3}$. (d) Eigenstate $|\psi_{1,(\frac{3}{4},\frac{1}{2})}\rangle$ for $\frac{\lambda}{2\pi}=\frac{2}{3}$.}
\end{figure}

We denote the quasienergy states by $|\psi_{b,\mathbf{k}}\rangle$ with $\mathbf{k}\equiv(k_X,k_P)$ and $b$ the band index counting from the bottom. To visualize the quasienergy states, we define the Husimi $Q$-function of a given eigenstate in phase space \cite{Knight2005}
\begin{equation}
Q_{b,{\mathbf k}}(\alpha,\alpha^*)\equiv\frac1\pi\langle\alpha|\psi_{b,{\mathbf k}}\rangle\langle\psi_{b,{\mathbf k}}|\alpha\rangle=\frac1\pi|\langle\alpha|\psi_{b,{\mathbf k}}\rangle|^2
\end{equation}
where $|\alpha\rangle$ is the coherent state introduced at the beginning in this section. In Figs.~\ref{fig:Qfunctions}(a) and (b), we plot the $Q$-functions of eigenstates $|\psi_{1,(0,0)}\rangle$ and $|\psi_{2,(0,0)}\rangle$ for $\frac{\lambda}{2\pi}=\frac{1}{2}$, which are the ground-like states  of the lower band and upper band, respectively. Comparing the $Q$-functions of the two states to the phase space lattices shown in Fig.~\ref{fig:psl}(a), we see that the $Q$-function of eigenstate $|\psi_{1,(0,0)}\rangle$ mostly occupies the negative phase space lattice $H_{sq}(X,P)<0$ while the $Q$-function of eigenstate $|\psi_{2,(0,0)}\rangle$ is shifted by $\pi$ along both $X$ and $P$ directions in phase space, mostly occupying the positive phase space lattice $H_{sq}(X,P)>0$. In fact, our system has a chiral symmetry defined by the chiral operator $\hat{T}_c\equiv e^{i\frac{\pi}{\lambda}\hat{X}}e^{i\frac{\pi}{\lambda}\hat{P}}$, i.e., $\hat{T}_cH_{sq}\hat{T}^\dagger_c=-H_{sq}$. Thus, for a given eigenstate $|\psi_{b,\mathbf{k}}\rangle$, there must be another eigenstate $\hat{T}^\dagger_c|\psi_{b,\mathbf{k}}\rangle$ with opposite quasienergy.
In Figs.~\ref{fig:Qfunctions}(c) and (d), we plot the $Q$-functions of eigenstates $|\psi_{1,(\frac{1}{4},\frac{1}{2})}\rangle$ and $|\psi_{1,(\frac{3}{4},\frac{1}{2})}\rangle$ for $\frac{\lambda}{2\pi}=\frac{2}{3}$, which are the degenerate states of the lower band shown in Fig.~\ref{fig:bands}(e). We see that the period along $X$-direction is $2\pi$ while the period along $P$-direction is $4\pi$. In fact, this degeneracy depends on the discrete translation operators we choose in Eq.~(\ref{eq:subgroup}). For $\frac{\lambda}{2\pi}=\frac{p}{q}$, the period of any $Q$-function of eigenstate is $2\pi p$-period in $P$-direction and $2\pi$-period in $X$-direction. Theses $p$-degenerate states are given by $$|\psi_{b,\mathbf{k}}\rangle, \ \hat{T}_2|\psi_{b,\mathbf{k}}\rangle,\  \hat{T}^2_2|\psi_{b,\mathbf{k}}\rangle,\ \cdot\cdot\cdot,\ \hat{T}^{p-1}_2|\psi_{b,\mathbf{k}}\rangle$$ with the same quasicoordinator but different quasimomenta. In the case of $\frac{\lambda}{2\pi}=\frac{2}{3}$, the two degenerate states of $|\psi_{b,\mathbf{k}}\rangle$ and $\hat{T}_2|\psi_{b,\mathbf{k}}\rangle$ in the $X$-representation are $\langle X|\psi_{b,\mathbf{k}}\rangle=\psi_{b,\mathbf{k}}(X)$ and $\langle X|\hat{T}_2|\psi_{b,\mathbf{k}}\rangle=e^{i\frac{2\pi}{\lambda} X}\psi_{b,\mathbf{k}}(X)$. Therefore, the $Q$-functions of the two degenerate states are the same in the $X$-dimension. But due to the relationship $\langle \psi_{b,\mathbf{k}} |\hat P|\psi_{b,\mathbf{k}}\rangle=\langle \hat{T}_2\psi_{b,\mathbf{k}} |\hat P|\hat{T}_2\psi_{b,\mathbf{k}}\rangle+2\pi$, the $Q$-functions are shifted by $2\pi$ in the $P$-dimension.

The underlying topology of a quasienergy band is defined by the Chern number \cite{PRA950221182017}
\begin{equation}\label{eq:ChernNumber}
c_b=\oint_\mathcal{C}\langle\psi_{b,\mathbf{k}}|\partial_\mathbf{k}|\psi_{b,\mathbf{k}}\rangle\cdot d\mathbf{k},
\end{equation}
where the contour $\mathcal{C}$ is integrated over the boundary of the Brillouin zone. The Chern number associated with a gap is subtle here. For the equilibrium systems, the Chern number of a gap is defined by the sum of the Chern numbers of the energy bands below the gap. However, in our present work, we are dealing with a Floquet system far from equilibrium. The general statistic law of the Floquet states for the long-time stationary state is an on-going research topic \cite{PRE790511292009,PRE820211142010,NJP180530082016}. Actually, the positive and negative sublattices shown in Fig.~\ref{fig:psl}(a) make no difference in the frame of Floquet theory. We assume that the statistic mechanics near the ground state of each sublattice can be described by an effective Floquet-Gibbs states \cite{NJP180530082016}.  Therefore, we define the Chern number of a gap below (above) the zero energy line as the sum of Chern numbers of all the quasienergy bands below (above) the gap. As shown in Fig.~\ref{fig:bands}(a), the Chern number of some gaps are calculated and labelled symmetrically with respect to zero energy line.

\subsection{Full Dissipative Quantum Dynamics}\label{sec:Dissipative}

\begin{figure*}
\centering
\includegraphics[width=\linewidth]{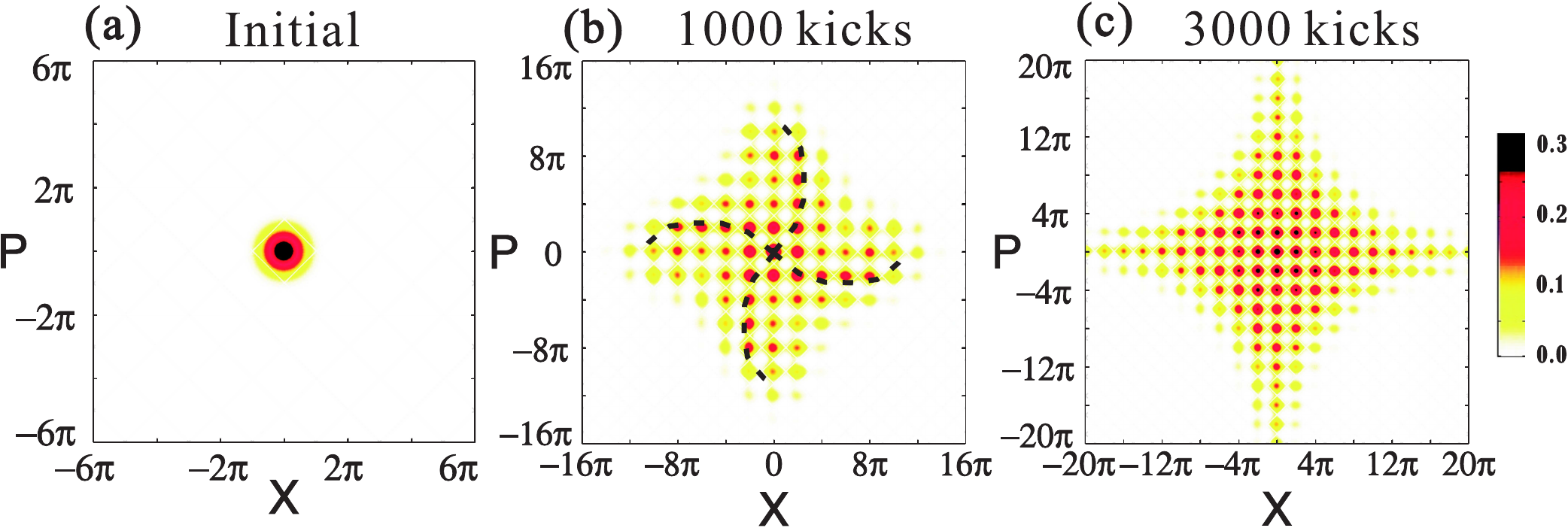}
\caption{\label{fig:disQs}{\bf Dissipative quantum dynamics.} (a) Husimi $Q$-function of the initial state, which is the ground state of the harmonic trapping potential. (b) Husimi $Q$-function after $1000$ kicks. We mark the main diffusion path by the dashed lines showing a chiral feature. (c) Husimi $Q$-function after $3000$ kicks. In all figures, we set kicking strength $K=0.1$, dissipation rate $\kappa=0.0001$. }
\end{figure*}

In the above section, our analysis is based on the rotating wave approximation where the kicking strength needs to be weak $|K|\ll 1$. In this section, we will investigate the full quantum dynamics of KHO based on the original full Hamiltonian (\ref{eq:dimensionlessKHO}) and confirm the validity of the rotating wave approximation, which is used to derive the effective Hamiltonian (\ref{eq:Hamiltoniansq}). From a practical point of view, the oscillators are inevitably in contact with the environment, which is conventionally modeled by a harmonic bath model. The coupling with the environment results in dissipation or decoherence of the quantum system. Here, we describe the dissipative dynamics of the quantum KHO by the following master equation,
\begin{equation}\label{eq:MasterEquation}
\frac{d{\rho}}{dt}=-\frac{i}{\lambda}[H_s(t),\rho]+\kappa(n_0+1)\mathcal{D}[\hat{a}]{\rho}+\kappa n_0\mathcal{D}[\hat{a}^\dagger]{\rho},
\end{equation}
where $\kappa$ characterizes the dissipation rate and $n_0$ is the Bose-Einstein distribution of the thermal bath. The dissipative dynamics is described by the Lindblad superoperator defined by $\mathcal{D}[\hat O]\rho\equiv\hat{O}\rho\hat{O}^\dagger-\frac12(\hat{O}^\dagger\hat{O}\rho+\rho\hat{O}^\dagger\hat{O})
$, where $\hat{O}$ is an arbitrary operator. The two Lindblad terms in Eq.~(\ref{eq:MasterEquation}) represents relaxation and heating processes respectively. We notice that some authors also choose the non-Lindblad Caldeira-Leggett master equation to describe the dissipative dynamics \cite{PRA950221182017}. Here, we choose the Lindblad master equation (\ref{eq:MasterEquation}) since it can give the correct thermal equilibrium state of harmonic oscillator without kicking force while the non-Lindblad Caldeira-Leggett master equation cannot \cite{Ramazanoglu2009}.

\begin{figure}
\centering
\includegraphics[width=\linewidth]{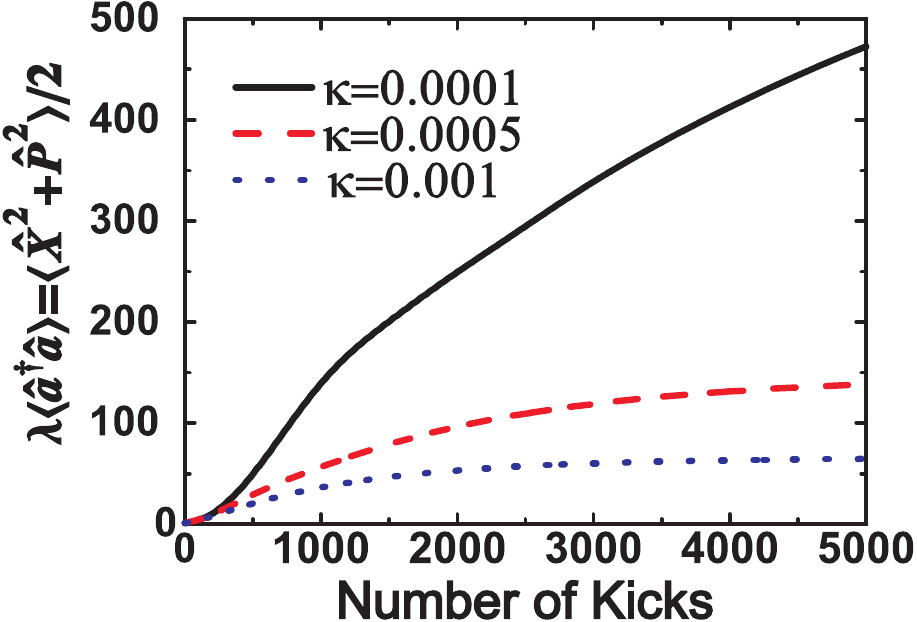}
\caption{\label{fig:dissipation}{\bf The average energy versus kick number.}  The averaged energy $\lambda\langle \hat{a}^\dagger \hat{a} \rangle=\langle \hat{X}^2+\hat{P}^2 \rangle/2$ is proportional to the area of the lattice size shown in Fig.~\ref{fig:disQs}. The black solid line, red dashed line and blue dotted line correspond to the dissipation rate $\kappa=0.0001$, $\kappa=0.0005$ and $\kappa=0.001$ respectively. The kicking strength is $K=0.1$. }
\end{figure}

As the kicks act as delta-functions, we can separate the dissipative dynamics from the kicking dynamics. In order to solve the dissipative dynamics, we define the characteristic function of the Wigner distribution by \cite{Knight2005}
$w(s,k)\equiv\int dx\,e^{ixk/\lambda}\langle x+\frac{s}{2}|\rho|x-\frac{s}{2}\rangle,
$
Then the master equation (\ref{eq:MasterEquation}) without kicking can be transformed into the following Fokker-Planck equation \cite{PRA950221182017}
\begin{equation}
\partial_tw+(\frac \kappa2-\lambda k)\partial_sw+(\frac \kappa2+\lambda s)\partial_kw=-\frac{\kappa}{2\lambda}(n_0+\frac 12)(s^2+k^2)w.
\end{equation}
The dissipative dynamics between two successive kicks is solvable from the above Fokker-Planck equation. Given the initial state at the moment right after $n-1$ kicks $w(s,k;\tau_{n-1}^+)$, where $\tau_{n-1}^+=(n-1)\tau+\Delta$ with a positive infinitesimal increment $\Delta$, the final state at the moment right before $n$ kicks $w(s,k;\tau_n^-)$ with $\tau_n^-=n\tau-\Delta$, is given by the following map \cite{PRA950221182017}
\begin{equation}\label{mapdissipative}
w(s,k;\tau_n^-)=e^{-\frac{n_0+1/2}{2\lambda}(1-e^{-\kappa\tau})(s^2+k^2)}w(s_r,k_r;\tau_{n-1}^+)
\end{equation}
with $s_r=e^{-\frac{\kappa\tau}{2}}(k\sin\tau+s\cos\tau)$ and $k_r=e^{-\frac{\kappa\tau}{2}}(k\cos\tau-s\sin\tau)$. The kicking dynamics is an instantaneous unitary transformation $\rho\rightarrow{\rho'}=\hat{U}_K\rho\hat{U}_K^\dagger$ with $\hat{U}_K\equiv e^{-iK\tau/\lambda\cos \hat{X}}$. In Appendix~\ref{app:kickingdynamics}, we prove that the corresponding map of the characteristic function of the Wigner distribution at the time $\tau_n=n\tau$ is given by
\begin{equation}\label{mapkick}
w(s,k;\tau_n^+)=\sum_{j=-\infty}^{\infty}J_j(2K\sin\frac s2)w(s,k+j\lambda;\tau_n^-),
\end{equation}
where the $J_j$ are the $j$th-order cylindrical Bessel function. Hence, the full dynamics of the quantum KHO in contact with a thermal bath is realized by applying the two maps (\ref{mapdissipative}) and (\ref{mapkick}) sequentially. From the characteristic function $w(s,k)$, it is direct to obtain the corresponding Husimi-$Q$ function (see Appendix~\ref{app:husimiQ}).

In Fig.~\ref{fig:disQs}(a), we evolve the dynamics of the system starting from the ground state of the harmonic oscillator. We then plot the Husimi $Q$-functions of the states after $1000$ and $3000$ kicks in Fig.~\ref{fig:disQs}(b) and Fig.~\ref{fig:disQs}(c) respectively. We see clearly that a final state with square lattice structure in phase space forms gradually revealing the underlying square structure of Hamiltonian (\ref{eq:Hamiltoniansq}).
Interestingly, we find that the transient state shown in Fig.~\ref{fig:disQs}(b) has no reflection symmetries with respect to $X$ and $P$ although the RWA Hamiltonian (\ref{eq:Hamiltoniansq}) has. There is a chiral feature as marked by the dashed lines along the backbone of the quasiprobability distribution. This chirality is a reflection of the topological property of our system and the noncommutative geometry \cite{Connest1994,JOMP1994} of the phase space. As approaching the stationary state, the chirality disappears in the end.

\begin{figure}
\centering
\includegraphics[width=\linewidth]{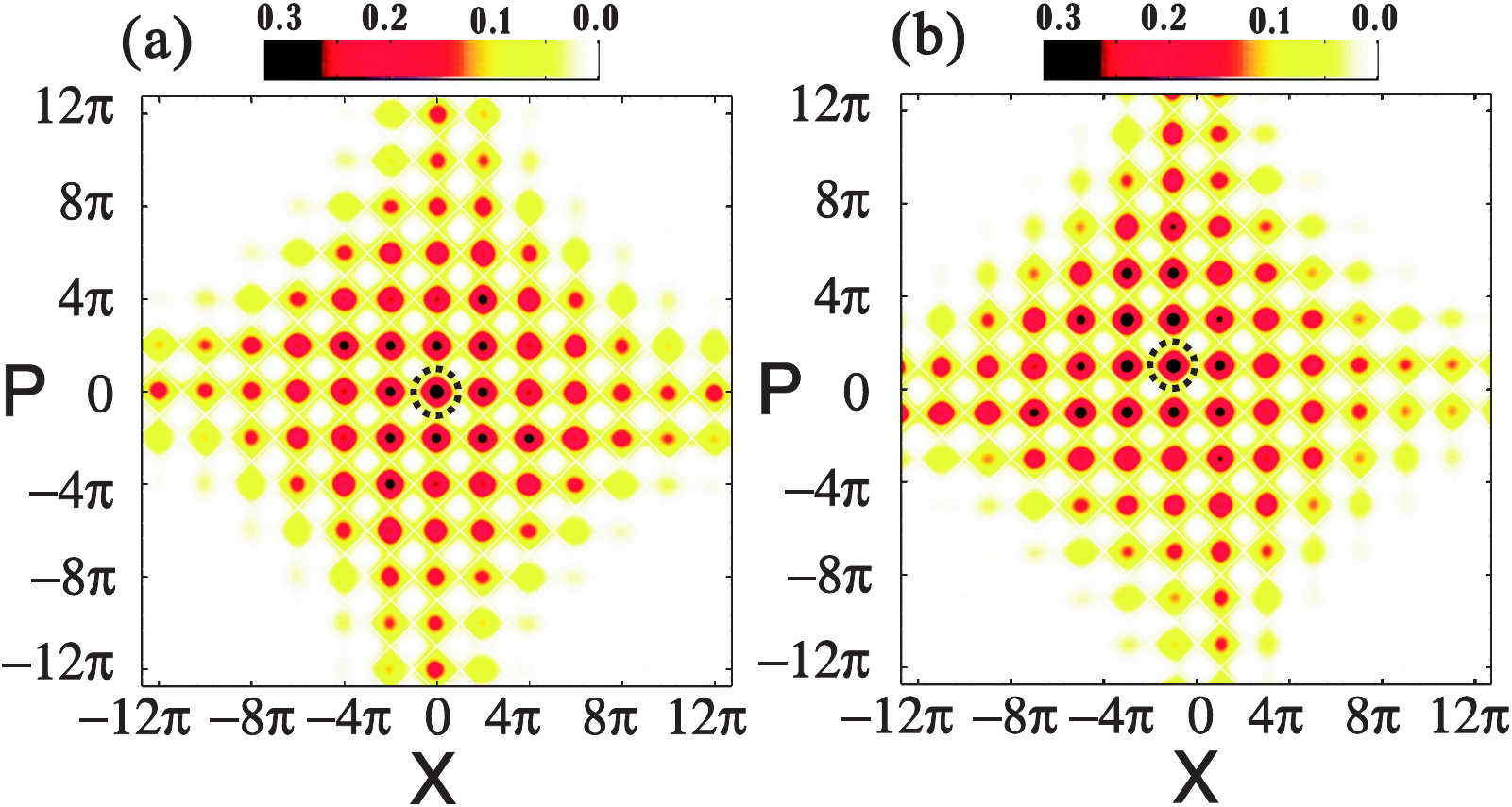}
\caption{{\label{fig:Q2Initials}\bf Husimi $Q$-functions for different initial states.} The initial state is a coherent state with its center locating on $(0,0)$ and $(-\pi,\pi)$ as marked by two dashed circles in figures (a) and (b) respectively. The $Q$-functions are plotted after $3000$ kicks. Other parameters: $K=0.1$, $\kappa=0.0001$. }
\end{figure}

Without dissipation, the quantum KHO will experience unbounded diffusion for resonant condition, where the average energy of the harmonic oscillator increases infinitely due to the energy pump from kicking \cite{PhysRevA-80-023414-2009}. When dissipation is present, the diffusion process approaches a nonequilibrium stationary state with a finite size in phase space depending on the driving strength and dissipation rate. In Fig.~\ref{fig:dissipation}, we plot the average energy of the KHO $\lambda\langle \hat{a}^\dagger \hat{a} \rangle=\langle \hat{X}^2+\hat{P}^2 \rangle/2$ as a function of the kicking number for different dissipation rates. We see that the smaller the dissipation rate is, the lager the phase space lattice is in the long-time limit. If the dissipation rate is so strong that the system can relax to its ground state during the successive kicks, the lattice state cannot be formed in phase space. Therefore, in order to create a phase space lattice with enough large size, the dissipation rate has to be much weaker than the kicking strength, i.e., $\kappa\ll |K|$.

In Figs.~\ref{fig:psl}(a) and (b), we also notice that there are actually two identical square lattices with a relative shift in phase space, which support eigenstates with positive and negative quasienergy respectively.
In Figs.~\ref{fig:Q2Initials}(a) and (b), we plot the two Husimi $Q$-functions evolving from two coherent states with different initial positions in phase space, i.e., $(X,P)=(0,0)$ and $(X,P)=(-\pi,\pi)$ respectively. We see that a state initially prepared on one sublattice stays on that lattice during the evolution and has negligible occupation on the other sublattice. This is different from the static potential, where the minimum points correspond to stable state while maximum points correspond to unstable state. Since we are working on a dynamical system far from equilibrium, both minimum and maximum points of the Hamiltonian in phase space are stable; only the saddle points are unstable. This is the reason why we define the Chern number of the gaps symmetrically with respect to the zero line for the Hofstadter's spectrum in Fig.~\ref{fig:bands}(a).

\section{Many-body Dynamics}\label{sec:Manybody}

In the above discussion, we have neglected the interaction terms in the original Hamiltonian (\ref{ManybodyH}). From this section, we will consider the interactions between particles. Using the free time-evolution operator $\hat{O}(t)=\exp(i\lambda\sum_{i}\hat{a}_i^\dagger
\hat{a}_it)$ defined at the beginning in Sec.~\ref{sec:PSCrystals}, the total Hamiltonian in the rotating frame is given by the canonical transformation, i.e.,
$\hat{O}(t)H(t)\hat{O}^\dagger(t)-iO(t)\dot{O}^\dagger(t)$. In the RWA, we drop the fast oscillating
terms and arrive at the time-independent Hamiltonian
\begin{eqnarray}\label{RWAH}
H_{RWA}^T &=&\sum_{i} H_{RWA}(\hat{X}_i,\hat{P}_i)+ \ \sum_{i<j}U(\hat{X}_i,\hat{P}_i;\hat{X}_j,\hat{P}_j).
\end{eqnarray}
Here, $H_{RWA}(\hat{X}_i,\hat{P}_i)$ is the single-particle RWA Hamiltonian given by Eq.~(\ref{eq:RWAHamiltonian}). The RWA interaction potential $U(\hat{X}_i,\hat{P}_i;\hat{X}_j,\hat{P}_j)$ is the
time-independent part of transformed real space interaction potential
$\hat{O}(t)V(\hat{x}_i-\hat{x}_j)\hat{O}^\dagger(t)$. In general,
$U(\hat{X}_i,\hat{P}_i;\hat{X}_j,\hat{P}_j)$ is defined in phase space and depends on both coordinates and
momenta of two particles. Thus, we call $U(\hat{X}_i,\hat{P}_i;\hat{X}_j,\hat{P}_j)$ the \textit{phase space interaction potential}. We aim to determine the explicit form of $U(\hat{X}_i,\hat{P}_i;\hat{X}_j,\hat{P}_j)$ in this section.

\subsection{Phase Space Interaction Potential }\label{sec:PhaseSpaceInteraction}

For two arbitrary particles, we introduce the operators $
\hat{X}_c\equiv(\hat{X}_1+\hat{X}_2)/2, \  \hat{P}_c\equiv(\hat{P}_1+\hat{P}_2)/2
$ representing the coordinator and momentum of two particles' center of mass, and the operators $\Delta \hat{X}\equiv\hat{X}_1-\hat{X}_2$, $\Delta
\hat{P}\equiv\hat{P}_1-\hat{P}_2$ representing their relative displacement in phase space.
We further define the operator of \textit{phase space
distance} by
\begin{eqnarray}
\hat{R}\equiv\sqrt{\Delta
\hat{X}^2+\Delta \hat{P}^2}.
\end{eqnarray}
It is important to notice that the background of the phase space interaction potential is a noncommutative space. From the commutation relationship $[\hat{X}_i,\hat{P}_j]=i\lambda\delta_{ij}$, we have
$[\Delta \hat{X},\Delta \hat{P}]=i(2\lambda)$, and $[\hat{R}^2,\hat{X}_c]=0$ which means the motion of two particles' center
of mass and their relative motion are independent. Thus, we write the common eigenstate of commutative operators $\hat{R}^2$ and $\hat{X}_c$ as a product state
$
\Psi(X_1,X_2)= f(X_c)\Phi(X_1-X_2),
$
where the wave function $f(X_c)$
is the state of two particles' center of mass and the wave function
$\Phi(X_1-X_2)$ describes their relative motion.
Reminiscent of the Hamiltonian operator of a harmonic
oscillator, the eigenvalues of operator $\hat{R}^2$ are
given by $4\lambda(N+1/2)$ with $N=0,1,2,\cdot\cdot\cdot$. Therefore, the eigenvalues
of the operator $\hat{R}$ are given by 
\begin{eqnarray}\label{QuantizedRN}
R_N=2\sqrt{\lambda(N+\frac{1}{2})},\ \ \  N=0,1,2,\cdot\cdot\cdot.
\end{eqnarray}
For each $N$, the corresponding eigenstate is given by
\begin{eqnarray}\label{QuantizedPhiN}
\Phi_N(\Delta X) =\Big(\frac{1}{\sqrt{2\lambda\pi} 2^N
N!}\Big)^{\frac{1}{2}}H_N\Big(\sqrt{\frac{1}{2\lambda}} \Delta
X\Big)e^{-\frac{1}{4\lambda}\Delta X^2},
\end{eqnarray}
where
$H_N(\bullet)$ is the Hermite polynomial of degree $N$. We choose
functions $\delta(X_c-C)$, i.e., the eigenstate of operator
$\hat{X}_c$, as the basis of two particles' center of mass. Therefore, we use the Dirac notation $|N,C\rangle$ to represent the
total eigenstate, which is determined by two good quantum numbers $C$ and
$R_N$, i.e., $\hat{X}_c|N,C\rangle=C|N,C\rangle$ and
$\hat{R}|N,C\rangle=R_N|N,C\rangle$. In the coordinate representation, the total eigenstate has the explicit form $\langle X_1,X_2|N,C\rangle=\Psi_{N,C}(X_1,X_2) =\delta(X_c-C)\Phi_N(\Delta X)$.

There is a fundamental difference between the commutative real space and the noncommutative phase space. The concept of \textit{point} is meaningless in noncommutative space. Instead, we are only allowed to define the coherent state $|\alpha\rangle$ as the \textit{point} in noncommutative geometry. Similarly, the concept of \textit{distance} also needs to be reexamined.  The distance of two particles in real space is a continuous
variable from zero to infinity. However, the distance in phase space is a quantized variable
and has a lower limit $\sqrt{2\lambda}$ as seen from Eq.~(\ref{QuantizedRN}). Here, we actually provide a description for the quantization of the noncommutative background.


We now start to determine the phase space interaction potential $U(\hat{X}_1,\hat{P}_1;\hat{X}_2,\hat{P}_2)$. From the transformation (\ref{xpXP}), the relative displacement of two particles in
the rotating frame is
$
 \hat{O}(t)(\hat{x}_1-\hat{x}_2)\hat{O}^\dagger(t) =
 \Delta\hat{P}\sin t+\Delta\hat{X}\cos t.
$
Therefore, for a given real space interaction potential, we have $\hat{O}V(\hat{x}_1-\hat{x}_2)\hat{O}^\dagger=\int_{-\infty}^{+\infty}dq V_q
\hat{Q}$ with the Fourier coefficients $V_q=\frac{1}{2\pi}\int_{-\infty}^{+\infty}dxV(x)e^{-iqx}$ and the operator $\hat{Q}\equiv
\exp \Big[iq(\Delta\hat{P}\sin t+\Delta\hat{X}\cos t)\Big].$
The matrix element of the operator $\hat{Q}$ in the $\hat{R}$-representation is
given by the Laguerre polynomials \cite{QuantumOpt33591991,NewJPhys-18-0230065-2016} 
\begin{eqnarray}\label{MLaguerre}
\langle N|\hat{Q}|M\rangle =e^{-\frac{\lambda
q^2}{2}+i(M-N)(\frac{\pi}{2}-\frac{\Omega}{k} t)}
\sqrt{\frac{N!}{M!}}\ (\lambda
q^2)^{\frac{M-N}{2}}L_N^{M-N}(\lambda q^2),\ \ \ \ \ \
\end{eqnarray}
where $|N\rangle$ and $|M\rangle$ are the eigenstates of the operator $\hat{R}$ given by Eq.~(\ref{QuantizedPhiN}). In the RWA, we only keep the time-independent diagonal elements
of the matrix (\ref{MLaguerre}), i.e., $\langle N|\hat{Q}|N\rangle$ with $N=0,1,2,\cdot\cdot\cdot$. Thus, given an arbitrary real space interaction potential $V(x_1-x_2)$, we find a compact
expression for the phase space interaction potential
\begin{eqnarray}\label{RWA-VLa}
U(\hat{R})=\int_{-\infty}^{+\infty}dq V_q e^{-\frac{\lambda
q^2}{2}} L_{\frac{1}{4\lambda}\hat{R}^2-\frac{1}{2}}(\lambda q^2).
\end{eqnarray}
In the eigenbasis $|N,C\rangle$, we have $$U(\hat{R})=\sum_N U(R_N)\int dC  |N,C\rangle\langle N,C|.$$ Here, the interaction potential
$U(\hat{R})$ takes the value $U(R_N)$ with $R_N=2\sqrt{\lambda(N+1/2)}$.



If the two paricles have spins, their spatial state is either antisymmetric or symmetric depending on the symmetry of total spin state, i.e., $$\psi_{\pm}(X_1,X_2)=\frac{1}{\sqrt{2}}\Big[\varphi(X_1)\phi(X_2)\pm\phi(X_1)\varphi(X_2)\Big].$$
We sperate the average phase space interaction potential by $\langle U\rangle_{\pm}=\langle
\psi_{\pm}(X_1,X_2)|U(\hat{R})|\psi_{\pm}(X_1,X_2)\rangle
\equiv U_c\pm U_e.$
Here, we have defined the direct interaction $U_c$ and the the exchange interaction $U_e$, respectively,
\begin{eqnarray}\label{}
\left\{
\begin{array}{lll}
U_c&\equiv&\langle\varphi(X_1)\phi(X_2)|U(\hat{R})|\varphi(X_1)\phi(X_2)\rangle\\
U_e&\equiv&\langle\varphi(X_1)\phi(X_2)|U(\hat{R})|\phi(X_1)\varphi(X_2)\rangle.\\
\end{array}
\right.
\end{eqnarray}
The direct interaction part $U_c=\frac{1}{2}(\langle U\rangle_++\langle
U\rangle_-)$ corresponds to the classical interaction while the exchange interaction part $U_e=\frac{1}{2}(\langle U\rangle_+-\langle
U\rangle_-)$ is a
pure quantum effect without classical counterpart, which we call $U_e$ the \textit{Floquet exchange interaction} for our system. In the $\hat{R}$-representation, they have been calculated in the Appendix \ref{app:UcUe}, i.e.,
\begin{eqnarray}\label{DQ}
U_c=\sum_NU(R_N) I_N, \ \ \ \
U_e=\sum_N(-1)^NU(R_N) I_N
\end{eqnarray}
with the overlap integral
\begin{eqnarray}\label{overlapIN}
I_N= \int  dC \Big|\langle \varphi(C+\frac{1}{2}\Delta
X)\phi(C-\frac{1}{2}\Delta X)|\Phi_N(\Delta X)\rangle\Big|^2.
\end{eqnarray}
In the Appendix \ref{app:CSR}, we have given the overlap integral $I_N(R)$ for the two displaced coherent states $\varphi(X)=\Big(\frac{1}{\sqrt{\pi\lambda}}\Big)^{\frac{1}{2}}e^{-\frac{1}{2\lambda}(X-R/2)^2}$ and $\phi(X)=\Big(\frac{1}{\sqrt{\pi\lambda}}\Big)^{\frac{1}{2}}e^{-\frac{1}{2\lambda}(X+R/2)^2}$, where $R$ is the distance between the centers of two
coherent states in phase space. Below, we will calculate the analytical expressions of $U_c(R)$ and $U_e(R)$ for contact and hardcore interactions of ultracold atoms.

\subsection{Applications}\label{sec:Applications}
In this section, we apply our general theory of phase space interaction to
the special cases of contact interaction and hardcore interaction for ultracold atoms. We show that, in quasi-1D, the point-like contact
interaction in real space becomes a long-range Coulomb-like
interaction in phase space. In pure 1D, the hardcore interaction in real space produces a quark-like confinement interaction potential in phase space, which increases linearly with the phase space distance of two atoms.

\subsubsection{Contact Interaction}\label{sec:ContactInteraction}

\begin{figure}\label{fig:PSI}
\centering
\includegraphics[scale=1.0]{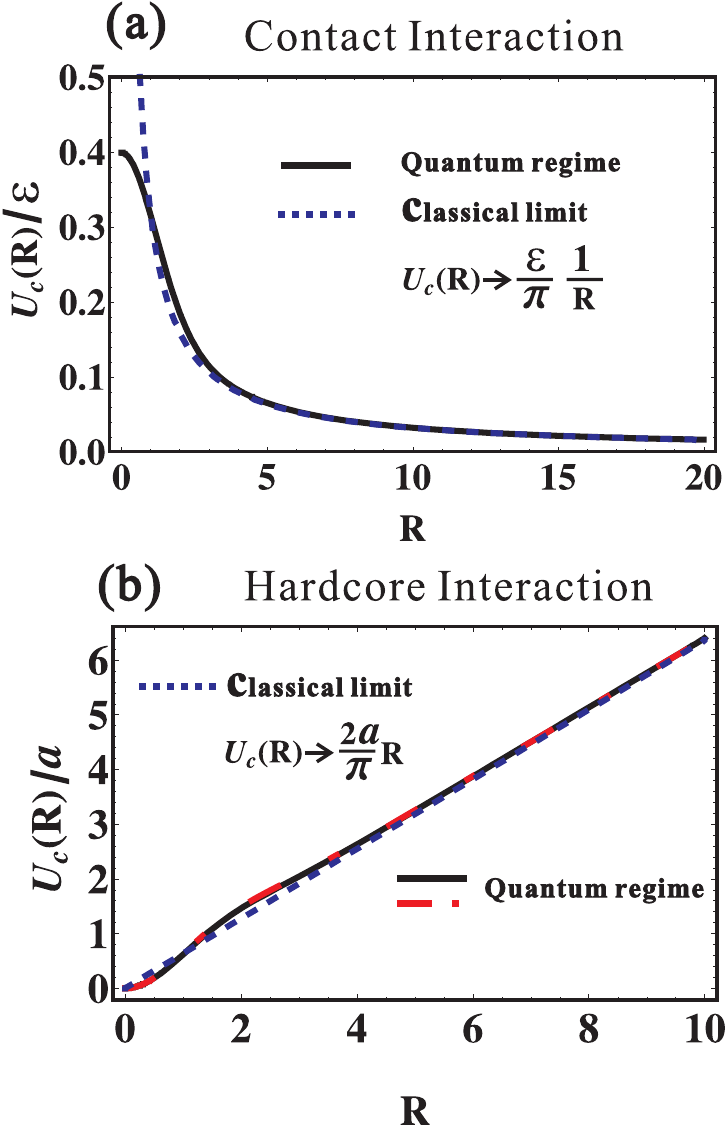}
\caption{\label{fig:PSIPs}{\bf Phase Space Interactions.}
(a) Contact interaction. The parameter $R$ is the distance between the centers of two  coherent states in phase space. The direct phase space interaction potential $U_c(R)$ is scaled by the strength of the contact interaction $\varepsilon$. The black solid curve is the result in quantum regime given by Eq.~(\ref{UcRcontact}) with $\lambda=1$. The blue dashed line is the Coulomb-type interaction (\ref{UcRcontactclassical}), which is the result in the classical limit. (b) Hardcore interaction. The direct phase space interaction potential $U_c(R)$ is scaled by the radius of hardcore interaction $a$. The black solid curve is the result in quantum regime given by Eq.~(\ref{UceHardcore}) with $\lambda=1$. The red dotted-dashed line is the result in the linear approximation given by Eq.~(\ref{hardcorepotential}) with $\lambda=1$. The blue dashed line is the linear interaction (\ref{linearpotential}) in the classical limit.
}
\end{figure}

%

In the experiments, the ultracold atoms are confined in one dimension if the transverse trapping frequency $\omega_{\perp}$
is much larger than the longitudinal trapping frequency $\omega_z$. If the characteristic length of transverse trapping $l_\perp\equiv\sqrt{\hbar/(m\omega_\perp)}$ is much larger than the cold atom's size, i.e., in the quasi-1D, the effective interaction between cold atoms is described by the contact
interaction $V(x_1-x_2)=\varepsilon\delta(x_1-x_2)$, where $\varepsilon$ is the interaction strength \cite{RevModPhys808852008,PRL819381998,PRL911632012003}. From $V_q=\frac{1}{2\pi}\int_{-\infty}^{+\infty}dx\varepsilon\delta(x)e^{-iqx}=\frac{\varepsilon}{2\pi}$ and Eq.~(\ref{RWA-VLa}), the phase space interaction potential can be calculated
\begin{eqnarray}\label{URNDelta}
U(R_N)=\varepsilon\frac{1+(-1)^N}{N\pi\sqrt{2\lambda}}\frac{\Gamma(\frac{N+1}{2})}{\Gamma(\frac{N}{2})}.
\end{eqnarray}
Here, $N=0,1,2,3\cdot\cdot\cdot$ and $\Gamma(\bullet)$ is the
gamma function. We see that $U(R_N)$ is zero for odd integer $N$
and finite for even integer $N$. The wave function of two atoms'
relative motion $\Phi_N(\Delta X)$ is antisymmetric for odd $N$,
which means the probability amplitude is zero when the two atoms contact each other.
The result is that the total average interaction of $\Phi_N(\Delta X)$ is zero for odd $N$.

The direct phase space interaction $U_c(R)$ and Floquet exchange interaction $U_e(R)$ of two cold atoms, which are described by two coherent states, can be calculated from
Eq.s~(\ref{DQ}), (\ref{URNDelta}) and (\ref{INSC})
\begin{eqnarray}\label{UcRcontact}
\left\{
\begin{array}{lll}
U_c(R)=\frac{\varepsilon}{\sqrt{2\pi\lambda}}\exp\Big(-\frac{R^2}{4\lambda}\Big)\
I_0\Big(\frac{R^2}{4\lambda}\Big)\\
U_e(R)=U_c(R).\\
\end{array}
\right.
\end{eqnarray}
Here, $R$ is the distance in phase space between the centers of two coherent states and $I_0(\bullet)$ is the zeroth order modified Bessel function
of the first kind. In the large distance limit, we use the asymptotic behavior of the special function
$I_0(z)\sim e^z/\sqrt{2\pi z}$ for $z\gg 1$ and have
\begin{eqnarray}\label{UcRcontactclassical}
U_c(R) &\sim&\frac{\varepsilon}{\pi }\frac{1}{R},\ \ \ \  \mathrm{for}
\ \ R\gg 2\sqrt{\lambda}.
\end{eqnarray}
In Fig.~\ref{fig:PSIPs}(a), we plot the $U_c(R)$ as function of $R$ and its
long-range asymptotic behavior. We see that a point-like contact
interaction indeed becomes a long-range Coulomb-like interaction in the long-distance limit, which is consistent with the pure classical analysis \cite{PRA-93-053616-2016}.

As shown in Eq.~(\ref{UcRcontact}), we also find that the Floquet exchange interaction is equal to the direct phase space interaction, i.e., $U_e(R)=U_c(R)$, and does not disappear even in the classical limit $\lambda\rightarrow 0$, which cannot happen in a static system. Usually, the effective spin-spin interaction in Heisenberg model comes from the quantum exchange interaction between the nearest-neighbouring electrons and cannot be explained by classical dynamics. One should always keep in mind that we are investigating the effective stroboscopic dynamics and the two atoms indeed collide with each other during every stroboscopic time step. The phase space interaction is actually the time-averaged real space interaction in one harmonic period. The spin-spin interaction in Heisenberg model is a short-range interaction due to the exponentially small wave function overlap of two next-nearest-neighbouring electrons. However, here in our system, the Floquet exchange interaction has long-range behavior following Coulomb's law. In the classical limit $\lambda\rightarrow 0$, the long-range Floquet exchange interaction can be viewed as an effective long-range spin-spin interaction induced by collision of two atoms. The equality $U_e(R)=U_c(R)$ comes from the $\delta$ function modelling the contact interaction and the fact that the spatial antisymmetric state of two atoms has zero probability to touch each other. If the interaction potential between cold atoms is different from the $\delta$-function model, it is possible to tune the phase space interaction $U_c(R)$ and collision-induced spin-spin interaction $U_e(R)$ independently in the experiments.

\subsubsection{Hardcore Interaction}\label{IVB}

If the characteristic length of transverse trapping $l_\perp$ is much smaller than the cold atom's size, which is called pure-1D, the contact interaction is no longer valid for the description of interaction between cold atoms \cite{PRL920304022004,Nature4292772004,Science30511252004,Science32512242009}. In this situation, the atom can be viewed as a hardcore particle with a radius $a$, which means the interaction potential between the two atoms is infinite when their distance is smaller than $2a$ and zero when the distance is larger than $2a$. Our theory of phase space interaction can be applied to the small hardcore limit $a\ll \sqrt{\lambda}$. Since the two atoms can not contact each other due to the hardcore interaction, the engenstates of phase space distance operator $\hat{R}$ have to be zero at zero distance, which means that only the odd eigenstates $\Phi_{2m+1}(\Delta X)$ with $m\in\mathbb{N}$ satisfy this condition. The even eigenstate should be reconstructed as $\Phi_{2m}(\Delta X)=\mathrm{sgn}(\Delta X)\Phi_{2m+1}(\Delta X)$ with $\mathrm{sgn}(\bullet)$ the sign function. The eigenstates $\Phi_{2m}(\Delta X)$ and $\Phi_{2m+1}(\Delta X)$ are degenerate with the same eigenvalue $R_{2m}=R_{2m+1}=2\sqrt{\lambda(2m+3/2)}$. In the Appendix \ref{app:hardcore}, we calculate the phase space interaction potential of the hardcore interaction potential for odd integers $N=2m+1$
\bea\label{UNhardcore}
U(R_N)&=&\frac{a\sqrt{2\lambda/\pi}}{2^{-N} N!}\sum_{k,l=0}^{[\frac N2][\frac N2]}\,\frac{(-1)^{k+l}(N!)^2(N-k-l)!}{2^{2k+2l}k!l!(N-2k)!(N-2l)!}\nonumber\\
&\approx&\frac{2a}{\pi}R_N.\ \ \ \ \
\eea
Here, $[\frac N2]$ means the closest integer number less than $\frac N2$. For even integers $N=2m$, we have $U(R_{2m})=U(R_{2m+1})$. Here, we find that $U(R_N)$ can be approximated very well by the linear relationship $U(R_N)\approx 2a\pi^{-1}R_N$.

The direct phase space interaction $U_c(R)$ and the Floquet exchange interaction $U_e(R)$ of two coherent states can be calculated from Eq.s~(\ref{DQ}), (\ref{UNhardcore})
\begin{eqnarray}\label{UceHardcore}
\left\{
\begin{array}{lll}
U_c(R)=2\sum_{m=0}U(R_{2m+1}) I_{2m+1}\\
U_e(R)=0.\\
\end{array}
\right.
\end{eqnarray}
Here, $R$ is phase space distance between the centers of two coherent states and the overlap integral $I_{2m+1}$ is given by Eq.~(\ref{INSC}). The zero Floquet exchange interaction comes from the degeneracy of the symmetric and antisymmetric states. Thus, there is no collision-induced spin-spin interaction for the hardcore interaction. Using the linear approximation $U(R_N)\approx2a\pi^{-1}R_N$ and Eq.~(\ref{UceHardcore}), we have
\begin{eqnarray}\label{hardcorepotential}
U_c(R)
&\approx&\frac{8a\sqrt{\lambda}}{\pi}\sum_{m=0}^\infty\,\frac{\sqrt{2m+3/2}}{(2m+1)!}\,\exp\Big(-\frac{R^2}{4\lambda}\Big)
\Big(\frac{R^2}{4\lambda}\Big)^{2m+1}.\ \ \
\end{eqnarray}
In the long-distance limit, we have the asymptotic expression of Eq.~(\ref{hardcorepotential}), i.e.,
 \begin{eqnarray}\label{linearpotential}
 U_c(R)\rightarrow \frac{2a}{\pi}R,\ \ \ \  \mathrm{for}
\ \ R\gg 2\sqrt{\lambda}.
 \end{eqnarray}
This is consistent again with the classical analysis \cite{PRA-93-053616-2016}. In Fig.~\ref{fig:PSIPs}(b), we plot the direct phase space interaction potential $U_c(R)$ as a functions of phase space distance $R$. We see that the linear relationship (\ref{linearpotential}) (blue dashed line) is a very good approximation of Eq.~(\ref{UceHardcore}) (black solid curve) and Eq.~(\ref{hardcorepotential}) (red dotted-dashed curve). It is interesting to find that the linear phase space interaction potential (\ref{linearpotential}) mimics the interaction potential between quarks in QCD \cite{PRD900740172014,PTEP083D022016}. Actually, this surprising behavior of hardcore atoms can be understood in a simple picture. Since the two atoms have a tiny hardcore radius $a$, they prefer to oscillate in a synchronized way, i.e., in phase. If the atoms are out of phase due to the finite phase space distance $R$, they are more likely to collide with each other during the oscillation. The collision effect becomes stronger as the phase space distance $R$ is larger, resulting in a confinement potential in the end.

\subsection{Classical Many-body Dynamics}\label{Sec:classicalmanybody}

\begin{figure*}
\centering
\includegraphics[scale=0.9]{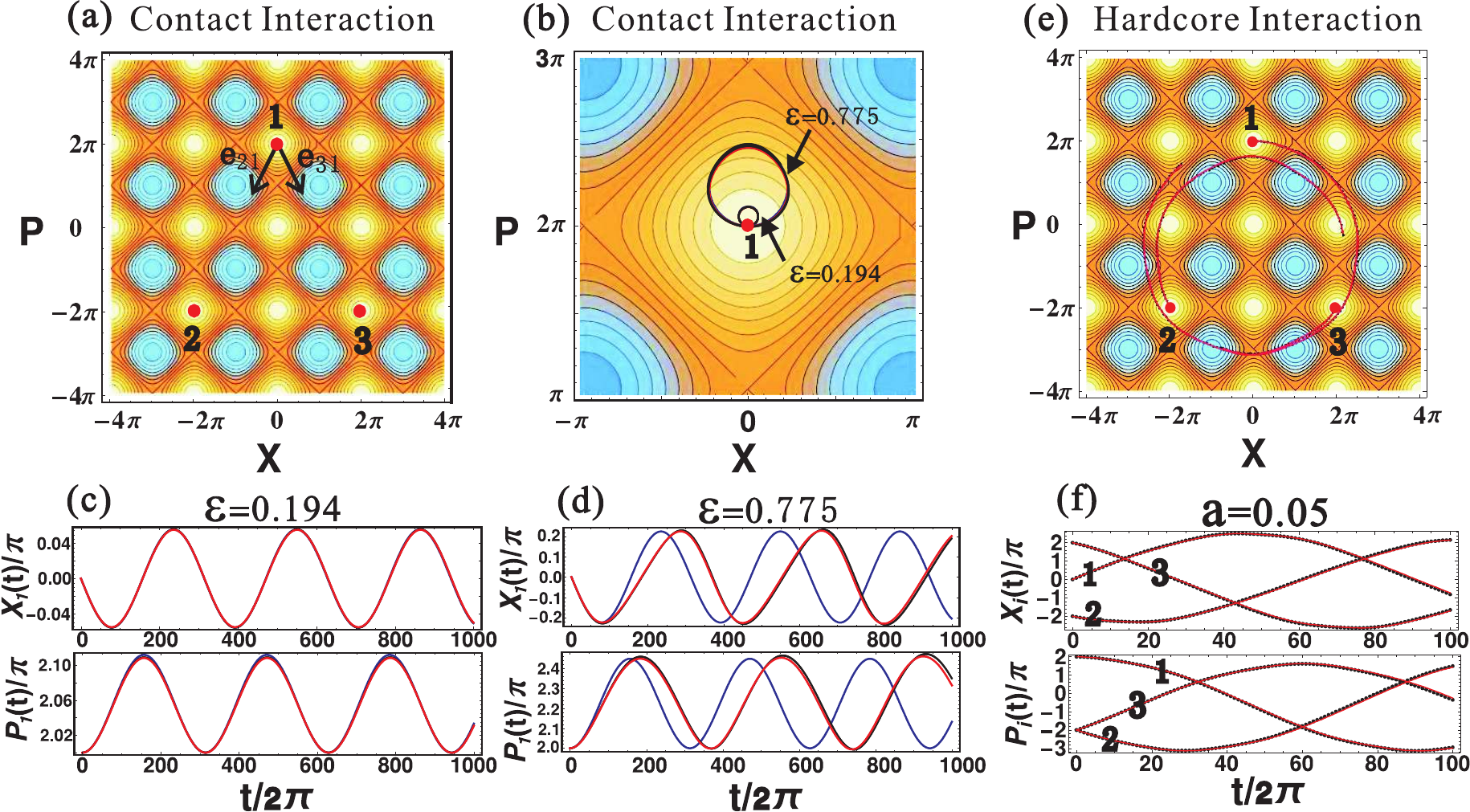}
\caption{\label{fig:3body}{\bf Classical Three-body Dynamics}. (a) The initial conditions of three atoms (red dots). (b) Phase space trajectories of the first atom with different contact interaction strengths $\varepsilon=0.194$ (small circle) and $\varepsilon=0.775$ (large circle). (c) Time evolutions of the first atom's position and momentum with contact interaction strength $\varepsilon=0.194$. (d) Time evolutions of the first atom's position and momentum with contact interaction strength $\varepsilon=0.775$. (e) Phase space trajectories of three atoms with hardcore interaction strength $a=0.05$. (f) Time evolutions of all the three atoms' positions and momenta with hardcore interaction strength $a=0.05$. In all figures, the black dots are the results obtained from Poincar\'e map and the red lines are the results calculated from the RWA EOM (\ref{MainEOMRWA}). In (b), (c) and (d), the blue lines are the results given by the linear solution (\ref{LinearEOM}) and the black dots are so dense that they look like lines. Other parameters: kicking strength $K=-0.02\pi^{-1}$. }
\end{figure*}

Although it is very difficult to numerically simulate the quantum many-body dynamics from the original many-body Hamiltonian (\ref{ManybodyH}), we can simulate the classical many-body dynamics and verify our theory of phase space interaction. From now on, we consider the classical dynamics of spinless atoms and replace all the operators by their corresponding classical quantities. The time evolution of the original coordinates, $x_i(t)$ and $p_i(t)$, of a single atom are given by the canonical equations
of motion (EOM) from Eq.~(\ref{ManybodyH})
\begin{eqnarray}\label{eomofxp}
\frac{dx_i}{dt}=\frac{\partial H(t)}{\partial p_i},\ \ \ \ \
\frac{dp_i}{dt}=-\frac{\partial H(t)}{\partial x_i}.
\end{eqnarray}
As seen from Eq.~(\ref{xpXP}), the values of
$X_i(t)$ and $P_i(t)$ can be obtained from the time evolution of
$x_i(t)$ and $p_i(t)$ stroboscopically every time period of
$2\pi$. In this sense, the $X_i(t)$ and
$P_i(t)$ define the time evolution of the amplitude and phase of an oscillating atom in the discrete time domain $t=2\pi m$ with
$m=0, 1, 2,\cdot\cdot\cdot$. This method is called
$Poincar\acute{e}\ map$ \cite{Xinbook,RLDevaney}.

In the rotating frame, we write the RWA many-body Hamiltonian explicitly for $q_0=4$
\begin{eqnarray}\label{ManybodyRWAH}
H_{RWA}^T &=&\sum_{i} \frac K2\Big(\cos X_i+\cos P_i\Big)+ \sum_{i<j}U_c(R_{ij}),
\end{eqnarray}
where $R_{ij}=\sqrt{(X_i-X_j)^2+(P_i-P_j)^2}$ is the classical phase space distance of two arbitrary atoms. Depending on the original interaction potential, the phase space interaction potential $U_c(R_{ij})$ takes the form of either Eq.~(\ref{UcRcontactclassical}) or Eq.~(\ref{linearpotential}). The EOM of $X_i(t)$ and $P_i(t)$ is described by
$
{dX_i}/{dt}={\partial H_{RWA}^T}/{\partial P_i}, \
{dP_i}/{dt}=-{\partial H_{RWA}^T}/{\partial X_i}.
$
Using Eq.~(\ref{ManybodyRWAH}), we have the explicit form of EOM
\begin{eqnarray}\label{MainEOMRWA}
\left\{
\begin{array}{lll}
\frac{d}{dt}X_i&=&-\frac{1}{2} K\sin P_i+\sum_j\frac{d U_c(R_{ij})}{dR_{ij}}\frac{P_i-P_j}{R_{ij}}
\\
\frac{d}{dt}P_i&=&\ \ \frac{1}{2} K\sin X_i-\sum_j\frac{d U_c(R_{ij})}{dR_{ij}}\frac{X_i-X_j}{R_{ij}} .
\end{array}
\right.
\end{eqnarray}
Using the above two methods, we can calculate the dynamics of many interacting atoms, and compare them to verify our phase space interaction theory.


In Fig.~\ref{fig:3body}, we investigate the dynamics of three interacting particles. For convenience, we introduce the complex position of the $j$th atom in phase space $\mathbf{Z}_j(t)\equiv X_j(t)+iP_j(t)$. As shown in Fig.~\ref{fig:3body}(a), we set the three atoms initially at the local equilibrium points of single particle Hamiltonian, i.e., $\mathbf{Z}_1(0)= 2\pi i$, $\mathbf{Z}_2(0)=-2\pi-2\pi i$ and $\mathbf{Z}_3(0)=2\pi-2\pi i$. If the displacement of each atom in phase space is small, i.e., $|\mathbf{Z}_i(t)-\mathbf{Z}_i(0)|\ll 1$, we can linearize the EOM (\ref{MainEOMRWA}), and have the following solution
\begin{eqnarray}\label{LinearEOM}
\mathbf{Z}_i(t)\approx\mathbf{Z}_i(0)+\frac{2}{K}\Big(e^{i\frac{1}{2}Kt}-1\Big)\sum_{j\neq i}\frac{d U_c}{dR_{ij}}\mathbf{e}_{ji},
\end{eqnarray}
where $\mathbf{e}_{ji}\equiv [\mathbf{Z}_j(0)-\mathbf{Z}_i(0)]/|\mathbf{Z}_j(0)-\mathbf{Z}_i(0)|$ is the unit vector directing from $i$th atom's initial position to $j$th atom's initial position in phase space as shown by the black arrows in Fig.~\ref{fig:3body}(a). The above linear solution indicates that each atom oscillates harmonically around a shifted equilibrium point $\mathbf{Z}_i(0)-2K^{-1}\sum_{j\neq i}({d U_c}/{dR_{ij}})\mathbf{e}_{ji}$ and with the amplitude $\Delta \mathbf{Z}_i =2K^{-1}|\sum_{j\neq i}({d U_c}/{dR_{ij}})\mathbf{e}_{ji}|$.

In Figs.~\ref{fig:3body}(b)-(d), we show the three-body dynamics with the real space contact interaction potential $V(x_i-x_j)=\varepsilon\delta(x_i-x_j)$, which is modeled by a Lorentz function $V(x_i-x_j)=\frac{\varepsilon}{\pi}\frac{\sigma}{(x_i-x_j)^2+\sigma^2}$ with $\sigma=0.1$ in our numerical simulations. The corresponding phase space interaction potential is given by $U_c(R_{ij})=\frac{\varepsilon}{\pi}\frac{1}{R_{ij}}$. We plot the phase space trajectories of the first atom for different interaction strengths $\varepsilon=0.194$ and $\varepsilon=0.775$ in Fig.~\ref{fig:3body}(b), and the corresponding time evolutions of positions and momenta in Figs.~\ref{fig:3body}(c) and (d). We see that the dynamics given by the three methods agree with each other very well for weak interaction $\varepsilon=0.194$ as shown in Fig.~\ref{fig:3body}(c). However, for a larger interaction $\varepsilon=0.775$, the linear solution (\ref{LinearEOM}) breaks down while the RWA EOM (\ref{MainEOMRWA}) is still a very good approximation as shown in Fig.~\ref{fig:3body}(d).

In Figs.~\ref{fig:3body}(e) and (f), we show the three-body dynamics with the hardcore interaction, which is modeled approximately by an inverse power-law potential $V(x_i-x_j)=\frac{(2a)^n}{(x_i-x_j)^n}$ with a high power $n=20$ in our numerical simulations. The corresponding phase space interaction potential is given by $U_c(R_{ij})=2a\pi^{-1}R_{ij}$. As shown in Fig.~\ref{fig:3body}(e), for a small hardcore radius $a=0.05$, the phase space interaction is already strong enough to make the three particles overcome the potential barrier of phase space lattice and exhibit global motions. In Fig.~\ref{fig:3body}(f), we compare the results from Poincar$\acute{e}$ map (black dots) and the RWA EOM (\ref{MainEOMRWA}) (red lines), which agree with each other very well.

\subsection{Dynamical Crystals}\label{sec:DynamicalCrystals}

\begin{figure}
\centering
\includegraphics[width=\linewidth]{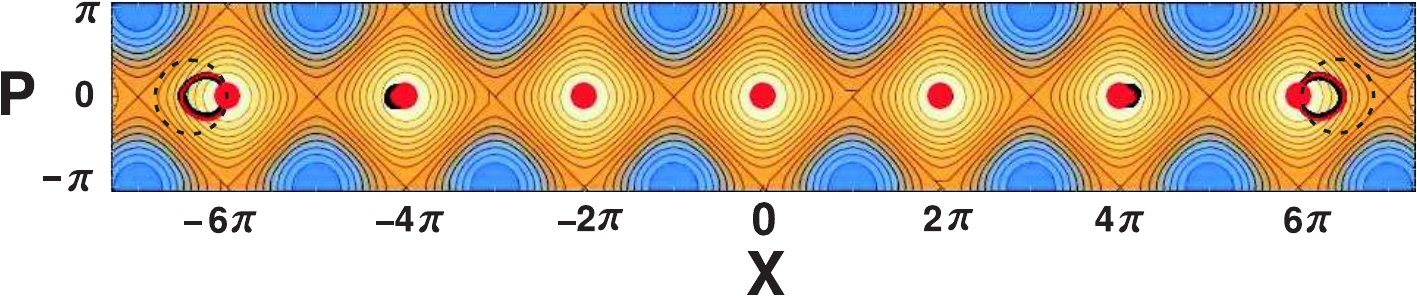}
\caption{{\label{fig:1DDC}\bf One-dimensional Dynamical Crystal.} The initial positions of seven atoms are marked by red dots. Black lines and red lines are the phase space trajectories calculated from Poincar\'e map and RWA EOM (\ref{MainEOMRWA}) with contact interaction strength $\varepsilon=0.194$. The one-dimensional dynamical crystal survives if the two atoms at the ends oscillate inside the dashed circles.  Kicking strength: $K=-0.02\pi^{-1}$.
}
\end{figure}

In Fig.~\ref{fig:1DDC}, we show the dynamics of seven interacting atoms for contact interaction with $\varepsilon=0.194$. The seven atoms are located initially at the equilibrium points with zero momenta as shown by the seven red dots. It can be seen that the two atoms at the ends oscillate with the largest amplitude. If the interaction is weak enough, the atoms only oscillate locally around their equilibrium points. If the interaction is strong enough, the two edge atoms can overcome the potential barrier of the phase space lattice, and destroy the crystal state. The existence of the crystal state is guaranteed by the condition that the oscillating amplitude of the edge atom $\Delta \mathbf{Z}_{\mathrm{edge}} $ is smaller than the radius of the dashed circle indicated in Fig.~\ref{fig:1DDC}, i.e., $\Delta \mathbf{Z}_{\mathrm{edge}} <(\sqrt{2}-1)\pi$. We can estimate the critical condition from the linear solution (\ref{LinearEOM}). For the contact interaction, the oscillating amplitude of the edge atom converges for infinite atoms, i.e., $\Delta \mathbf{Z}_{\mathrm{edge}}=\frac{\varepsilon}{12\pi K}$. Therefore, the critical interaction strength for the existence of 1D crystal state in phase space is given by
\begin{eqnarray}\label{DCCI}
\varepsilon_c\approx12(\sqrt{2}-1)\pi^2K.
\end{eqnarray}
For hardcore interaction with $U_c(R_{ij})=2a\pi^{-1}R_{ij}$, the oscillating amplitudes of the two edge atoms can be estimated by $\Delta \mathbf{Z}_{\mathrm{edge}}=\frac{4a}{\pi K}(N-1)$, where $N$ is the number of atoms. The oscillating amplitude $\Delta \mathbf{Z}_{\mathrm{edge}}$ increases linearly with the number of atoms, which means it is impossible to create an infinitely long 1D crystal state with hardcore interaction. For a given kicking strength $K$ and hardcore radius $a$, the critical atom number for the existence of 1D crystal state is
\begin{eqnarray}\label{DCHI}
N_c\approx(\sqrt{2}-1)\pi^2\frac{K}{4a}.
\end{eqnarray}
We call the stable crystal state in phase space formed by many atoms, the \textit{dynamical crystals}.

One should distinguish the concepts of phase space lattice discussed in Sec.~\ref{sec:PSCrystals} and the dynamical crystal introduced here. Phase space lattice refers to the periodic structure in phase space of the single-particle Hamiltonian (\ref{eq:RWAHamiltonian}) without consideration of atomic interaction, while the dynamical crystal refers to the many-body state formed by interacting atoms. In the experiments, the dynamical crystal can be realized by two basic steps: first, prepare the initial state of atoms via applying a very strong static optical lattice; then, suddenly turn off the strong static optical lattice and add a weak optical lattice stroboscopically.

If the atoms have spins and tightly bound at the their fixed points by the phase space lattice, the direct phase space interaction $U_c(R)$ does not play a role in the dynamics. However, as discussed in Sec.~\ref{sec:ContactInteraction}, the contact interaction can induce a Coulomb-like Floquet exchange interaction $U_e(R)=\frac{\varepsilon}{\pi }\frac{1}{R}$, and thus the system shown in Fig.~\ref{fig:1DDC} can be modelled by a 1D spin chains with isotropic spin-spin interaction. The famous Mermin-Wagner theorem claims that, at any nonzero temperature, a one- or two-dimensional isotropic Heisenberg model with finite-range exchange interaction can be neither ferromagnetic nor antiferromagnetic \cite{PRL1711331966}. This theorem clearly excludes a variety of types of long-range ordering in low dimensions, and is crucial to the search for low-dimensional magnetic materials in the recent years \cite{Nature5462652017,Nature5462702017,EPL101560032013}. Here in our model, the collision-induced spin-spin interaction has a Coulomb-like long-range behavior, which is beyond the definition of finite-range interaction in Mermin-Wagner theorem \cite{short-range}. Hence, the dynamical crystals actually provide a possible platform to test the Mermin-Wagner theorem and search for other new phenomena with long-range interactions such as causality and quantum criticality \cite{PRL1112072022013,Nature5111982014,PRL1130306022014,PRL1122014,PRL1141572012015,PRB931251282016,PRA930536202016}, nonlocal order \cite{PRL972604012006,PRB772451192008,Science3342002011}, etc.


\section{Summary}\label{Sec:SummaryandOutlook}

In summary, we have studied the possibility to create new physics by Floquet many-body engineering in the dynamical system of kicked interacting particles in 1D harmonic potential.
Our system exhibits surprisingly rich topological and many-body physics in 2D phase space. In the weak kicking strength regime $K\ll 1$, the single-particle RWA Hamiltonian has various lattice structures in phase space depending on the kicking period. The topological physics comes from the noncommutative geometry of phase space, which naturally provides a geometric quantum phase. We analyzed the topological quasienergy band structure of the square phase space lattice. We investigated the full dissipative quantum dynamics of a single kicked harmonic oscillator using master equation and the Fokker-Planck equation. The time evolution of the Husimi $Q$-functions confirms that the nonequilibrium stationary state is indeed a lattice state in phase space, but has a finite size due to the dissipation.

For the many-body dynamics, we made several findings and predictions based on the theory of phase space interaction potential. We found that the original contact interaction becomes a long-range Coulomb-like interaction in phase space, and the hardcore interaction becomes a quark-like confinement interaction in phase space. For the contact interaction, we predicted that the long-range Floquet exchange interaction does not disappear even in the classical limit, and can be viewed as collision-induced spin-spin interaction. We investigated the classical many-body dynamics and proposed the concept of dynamical crystals. We found that the contact interaction can create an infinitely long 1D dynamical crystal but the hardcore interaction cannot.

Finally, we point out that our method can increase the speed of numerical simulation significantly. For example, in simulating the dynamics of seven interacting atoms in Sec.~\ref{sec:DynamicalCrystals}, the method of Poincar\'e map based on the original Hamiltonian costs more than ten hours using \textit{Wolfram Mathematica} while the method based on the phase space interaction only needs one second. The reason is that our method only needs to calculate the dynamics on the stroboscopic time points by averaging the dynamics between stroboscopic steps using the phase space interaction potential.

Recently, we learned of a related study \cite{arXiv171010087} in which the authors also discussed that the contact interactions between atoms can result in exotic long-range interactions in the effective description of the resonantly driven many-body system.

\bigskip

\textbf{Acknowledgements}

LG and PL acknowledge financial support from Carl-Zeiss Stiftung (0563-2.8/508/2).

\appendix
\section{RWA Floquet Hamiltonian}\label{app:HRWA}

The method we adopt here is the same as that in \cite{NewJPhys-18-0230065-2016}. We start with the dimensionless Hamiltonian (\ref{eq:dimensionlessKHO}). To be clear, we write it again here
\begin{equation}\label{eq:dimensionlessKHOr}
 H_s=\frac12(\hat x^2+\hat p^2)+K\tau\cos\hat x\sum_{n}\,\delta(t-n\tau),
\end{equation}
where we have neglected the particle index and simplified the notation of the summation in the kicking part.
As discussed in the main text, we transform Eq.~(\ref{eq:dimensionlessKHOr}) into the rotating frame by employing the unitary transformation $\hat O=\exp\big[i(\hat a^\dagger \hat a+1/2)t\big]$ and using the relationship (\ref{xpXP})
\begin{eqnarray}\label{eq:HamiltonianRF}
H_{RF}&=&\hat OH_s\hat O^\dg-i\lambda\hat O\dot{\hat O}^\dg            \nn
&=&K\hat O\cos\hat x\,\hat O^\dg\sum_n\,\delta(\frac t\tau-n)          \nn
&=&\frac{K}{2}\Big[\hat M(\hat X,\hat P)+\hat M^\dg(\hat X,\hat P)\Big]\sum_n\,\delta(\frac t\tau-n),   \nn
\end{eqnarray}
where we define $\hat M(\hat X,\hat P)=\exp\Big\{i\big[\hat X\cos t+\hat P\sin t\big]\Big\}$. The harmonic term in Eq.~(\ref{eq:HamiltonianRF}) disappears due to the resonant condition.

The element of $\hat M(\hat X,\hat P)$ in the basis of Fock states is evaluated to be \cite{QuantumOpt33591991,NewJPhys-18-0230065-2016}
\begin{eqnarray}\label{eq:MOperator}
\langle l|\hat M(\hat X,\hat P)|k\rangle&=&e^{-\lambda/4+i(k-l)(\pi/2-t)}\sqrt{\frac{l!}{k!}}\left(\frac\lambda2\right)^{\frac{k-l}{2}}   \nn
&&\times L^{k-l}_l(\lambda/2).
\end{eqnarray}
Here the $L_l^{k-l}$ are the generalized Laguerre polynomials.
Inserting Eq.~(\ref{eq:MOperator}) into $\hat{H}_{RF}$ we have
\begin{eqnarray}\label{eq:HRF}
H_{RF}&=&\frac{K}{2}\bigg[\sum_{k,l}\,e^{-\lambda/4+i(k-l)(\pi/2-t)}
\sqrt{\frac{l!}{k!}}\left(\frac\lambda2\right)^{\frac{k-l}{2}}  \nn
&&\times L^{k-l}_l\left(\lambda/2\right)|l\rangle\langle k|+h.c.\bigg]\sum_n\,\delta(\frac t\tau-n).
\end{eqnarray}
The sum of the Dirac $\delta$-functions in Eq.~(\ref{eq:HRF}) obeys the following identity \cite{MZas2005}
\begin{equation}
\sum_n\,\delta(\frac t\tau-n)=\sum_n\,\cos\frac{2\pi nt}{\tau}.
\end{equation}
Making use of this relation and dropping all terms relevant to $t$ in $\hat{H}_{RF}$ (rotating wave approximation) we get
\begin{eqnarray}
H_{RWA}&=&\frac{K}{2}\bigg[\sum_{l,n}\,e^{-\lambda/4+inq_0\pi/2}\sqrt{\frac{l!}{(l+nq_0)!}}
\left(\frac\lambda2\right)^{\frac{nq_0}{2}}  \nn
&&\times L_{l}^{nq_0}\left(\lambda/2\right)|l\rangle\langle l+nq_0|+h.c.\bigg].
\end{eqnarray}
The sum over $n$ can be formulated as the sum over $k=l+nq_0$ with the help of the formula
\begin{eqnarray}
\frac{1}{q_0}\sum\limits_{j=1}^{q_0}\,e^{-i2\pi (k-l)j/q_0}=
\begin{cases}
0, &  (k-l)/q_0\not\in \mathbb{Z}   \nn
1, &  (k-l)/q_0\in \mathbb{Z},
\end{cases}
\end{eqnarray}
that is,
\begin{eqnarray}
H_{\it RWA}&=&\frac{K}{2q_0}\bigg[\sum_{j=1}^{q_0}\sum_{k,l}\,e^{-\lambda/4+i(k-l)(\pi/2-2\pi j/q_0)}
\sqrt{\frac{l!}{k!}}\left(\frac\lambda2\right)^{\frac{k-l}{2}} \nn
&&\times L^{k-l}_l\left(\lambda/2\right)|l\rangle\langle k|
+h.c.\bigg].
\end{eqnarray}
Using Eq.~(\ref{eq:MOperator}), we have the final effective Hamiltonian
\begin{eqnarray}
H_{RWA}&=&\frac{K}{2q_0}\sum_{j=1}^{q_0}\bigg[e^{i\hat{X}\cos(2\pi j/q_0)+i\hat{P}\sin(2\pi j/q_0)}+h.c.\bigg]  \nn
&=&\frac{K}{q_0}\sum_{j=1}^{q_0}\,\cos\left(\hat{X}\cos\frac{2\pi j}{q_0}+\hat{P}\sin\frac{2\pi j}{q_0}\right).
\end{eqnarray}


Another way to derive the effective Hamiltnoian Eq.~(\ref{eq:RWAHamiltonian}) is to start from the Floquet operator in one harmonic oscillation
\begin{equation}\label{eq:Fq0}
\hat{F}^{q_0}=\Big[e^{-i(\hat{x}^2+\hat{p}^2)\tau/2\lambda}e^{-iK\cos{\hat{x}}/\lambda}\Big]^{q_0}.
\end{equation}
Following the same procedure in \cite{PhysRevA-80-023414-2009} , we can reformulate Eq.~(\ref{eq:Fq0}) as
\begin{eqnarray}
\hat{F}^{q_0}=\prod_{j=0}^{q_0-1}\exp\Bigg\{-iK\cos\Big[\sqrt{\frac{\lambda}{2}}(\hat{a}e^{-i2\pi j/q_0}+\hat{a}^\dagger e^{i2\pi j/q_0})\Big]/\lambda\Bigg\}.\nonumber
\end{eqnarray}
Expanding $\hat{F}^{q_0}$ into a power series of the kicking strength $K$ and keeping the terms in the first order, we again get the effective Hamiltonian $\hat{H}_{RWA}$.

\section{Calculation of $\langle \alpha|H_{\it RWA}(\hat X,\hat P)|\alpha\rangle$}\label{app:Hdiag}

Defining displacement operator $D_\alpha\equiv e^{\alpha
\hat{a}^\dagger-\alpha^*\hat{a}}$, then we have the following relationship
\begin{eqnarray}\label{eq:DD}
\left\{
\begin{array}{lll}
D_\alpha D_\beta=e^{i\mathrm{Im}(\alpha\beta^*)}D_{\alpha+\beta},\\
D_\alpha
|\beta\rangle=e^{i\mathrm{Im}(\alpha\beta^*)}|\alpha+\beta\rangle.\\
\end{array}
\right.
\end{eqnarray}
We consider a general Hamiltonian $H=\cos(p\hat{X}+q\hat{P})$ which can be rewritten as
\begin{eqnarray}\label{}
H&=&\cos(p\hat{X}+q\hat{P})\nonumber\\
&=&\frac{1}{2}\Big(e^{ip\hat{X}+iq\hat{P}}+e^{-ip\hat{X}-iq\hat{P}}\Big)\nonumber\\
&=&\frac{1}{2}\Big[e^{-\sqrt{\frac{\lambda}{2}}(q-ip)\hat{a}^\dagger+\sqrt{\frac{\lambda}{2}}(q+ip)\hat{a}}+h.c.\Big]\nonumber\\
&=&\frac{1}{2}\Big[D_{-\sqrt{\frac{\lambda}{2}}(q-ip)}+D_{\sqrt{\frac{\lambda}{2}}(q-ip)}\Big].
\end{eqnarray}
Here we have used $\hat{X}=\sqrt{\frac{\lambda}{2}}(\hat{a}^\dagger+\hat{a})$ and $\hat{P}=i\sqrt{\frac{\lambda}{2}}(\hat{a}^\dagger-\hat{a})$.
Using the relationship (\ref{eq:DD}) and the identity $\langle\alpha|\beta\rangle=e^{-|\alpha|^2/2-|\beta|^2/2+\alpha^*\beta}$, we have the matrix element of Hamiltonian $H=\cos(p\hat{X}+q\hat{P})$ in coherent state representation
\begin{widetext}
\begin{eqnarray}\label{}
\langle \alpha|H|\beta\rangle&=&\frac{1}{2}\Big(\Big\langle
\alpha\Big|\beta-\sqrt{\frac{\lambda}{2}}(q-ip)\Big\rangle
e^{-i\sqrt{\frac{\lambda}{2}}\mathrm{Im}[(q-ip)\beta^*]}+\Big\langle
\alpha\Big|\beta+\sqrt{\frac{\lambda}{2}}(q-ip)\Big\rangle
e^{i\sqrt{\frac{\lambda}{2}}\mathrm{Im}[(q-ip)\beta^*]}\Big)\nonumber\\
&=&\frac{1}{2}\Big(e^{-\frac{1}{2}|\alpha|^2-\frac{1}{2}|\beta-\sqrt{\frac{\lambda}{2}}(q-ip)|^2
+\alpha^*\beta-\alpha^*\sqrt{\frac{\lambda}{2}}(q-ip)-i\sqrt{\frac{\lambda}{2}}\mathrm{Im}[(q-ip)\beta^*]}+
e^{(-\frac{1}{2}|\alpha|^2-\frac{1}{2}|\beta+\sqrt{\frac{\lambda}{2}}(q-ip)|^2
+\alpha^*\beta+\alpha^*\sqrt{\frac{\lambda}{2}}(q-ip)+i\sqrt{\frac{\lambda}{2}}\mathrm{Im}[(q-ip)\beta^*]}\Big)\nonumber\\
&=&
\frac{1}{2}e^{-\frac{1}{2}|\alpha|^2-\frac{1}{2}|\beta|^2+\alpha^*\beta-\frac{\lambda}{4}|q-ip|^2}
\Big(e^{\frac{\beta}{2}\sqrt{\frac{\lambda}{2}}(q+ip)+(\frac{\beta^*}{2}-\alpha^*)\sqrt{\frac{\lambda}{2}}(q-ip)
-\frac{1}{2}\sqrt{\frac{\lambda}{2}}[(q-ip)\beta^*-(q+ip)\beta]}+e^{-\frac{\beta}{2}\sqrt{\frac{\lambda}{2}}(q+ip)-(\frac{\beta^*}{2}-\alpha^*)\sqrt{\frac{\lambda}{2}}(q-ip)
+\frac{1}{2}\sqrt{\frac{\lambda}{2}}[(q-ip)\beta^*-(q+ip)\beta]}\Big)\nonumber\\
&=&
\frac{1}{2}e^{-\frac{1}{2}|\alpha|^2-\frac{1}{2}|\beta|^2+\alpha^*\beta-\frac{\lambda}{4}|q-ip|^2}
\Big(e^{\beta\sqrt{\frac{\lambda}{2}}(q+ip)-\alpha^*\sqrt{\frac{\lambda}{2}}(q-ip)}
+e^{-\beta\sqrt{\frac{\lambda}{2}}(q+ip)+\alpha^*\sqrt{\frac{\lambda}{2}}(q-ip)}\Big)\nonumber\\
&=&
\frac{1}{2}e^{-\frac{1}{2}|\alpha|^2-\frac{1}{2}|\beta|^2+\alpha^*\beta-\frac{\lambda}{4}|q-ip|^2}
\Big(e^{\sqrt{\frac{\lambda}{2}}(\beta-\alpha^*)q+i\sqrt{\frac{\lambda}{2}}(\beta+\alpha^*)p}
+e^{-\sqrt{\frac{\lambda}{2}}(\beta-\alpha^*)q-i\sqrt{\frac{\lambda}{2}}(\beta+\alpha^*)p}\Big).
\end{eqnarray}
\end{widetext}
By defining the average coordinator and momentum
\begin{eqnarray}\label{}
\left\{
\begin{array}{lll}
X\equiv\langle\alpha|X|\alpha\rangle=\sqrt{\frac{\lambda}{2}}(\alpha^*+\alpha)\\
P\equiv\langle
\alpha|P|\alpha\rangle=i\sqrt{\frac{\lambda}{2}}(\alpha^*-\alpha),\\
\end{array}
\right.
\end{eqnarray}
we have the diagonal elements of $H=\cos(p\hat{X}+q\hat{P})$
\begin{eqnarray}\label{eq:diagH}
\langle \alpha|H|\alpha\rangle =\exp\Big(-\frac{\lambda}{4}|q-ip|^2\Big)\cos(pX+qP).
\end{eqnarray}
Using Eq.~(\ref{eq:diagH}) by setting $p=\cos\frac{2\pi j}{q_0}$ and $q=\sin\frac{2\pi j}{q_0}$, we can easily obtain $\langle \alpha|H_{\it RWA}(\hat X,\hat P)|\alpha\rangle=e^{-\lambda/4}H_{\it RWA}(X,P)$.

\section{Zak's $kq$ Representation}\label{sec:appendixC}

The $kq$ representation introduced in Ref.~\cite{Zak1972} is the complete orthonormal basis constructed by the common eigenstates of the translation operators in both $X$ and $P$ directions. In general,
for the translation operation in $X$ direction $e^{i\hat{P}A/\lambda}$, the ``shortest" commutative translation operator in $P$ direction is $e^{i\hat{X}B/\lambda}$ with $B=2\pi\lambda/A$. Given the dimensionless Planck constant $\lambda=p/q$, where $p$ and $q$ are coprime integers, we choose $A=2\pi/q$ and $B=2\pi p$ for constructing our Zak's $kq$ representation. In the coordinate representation, the basis for given quantum numbers $k_X$ and $k_P$ is \cite{Zak1972}
\begin{eqnarray}\label{Zakbasis}
\phi_{{(k_X,k_P)}} (X)=\sqrt{\frac\lambda q}\sum_l e^{i k_X \frac{2\pi l}{q}}\delta(X-\lambda k_P - \frac{2\pi l}{q})
\end{eqnarray}
with $l\in \mathbb{Z}$. State function (\ref{Zakbasis}) is composed of a series of Dirac's delta function with shifted phases. Note that the quantum numbers $k_X$ and $k_P$ take value in the region $0\le k_X<q, 0\le k_P <1/p$, which is $q$-times larger than the Brillioun zone. Since function (\ref{Zakbasis}) is the eigenstate of translation operator $e^{i\frac{2\pi}{q\lambda}\hat{P}}$, it is also the eignestate of the translation operator $\hat{T}_X=\Big(e^{i\frac{2\pi}{q\lambda}\hat{P}}\Big)^q$,  defined by Eq.~(\ref{eq:subgroup}) in the main text, with the the same quasimomentum $k_X$ but different Brillioun range $0\le k_X<1$. States (\ref{Zakbasis}) with $q$ different quasimomenta, i.e., $\phi_{{(k_X,k_P)}}(X)$, $\phi_{(k_X+1,k_P)}(X)$, $\cdot\cdot\cdot$ and $\phi_{(k_X+q-1,k_P)}(X)$, can be treated as $q$ degenerate states of operator $\hat{T}_X$, which guarantee the completeness of the Zak's basis.

\section{Derivation of Eq.~(\ref{eq:butterfly})}\label{app:butterfly}

We outline how to derive Eq.~(\ref{eq:butterfly}) in the main text.
The quasienergy state $|\psi_{b,\mathbf{k}}\rangle$ can be spanned as $|\psi_{b,\mathbf{k}}\rangle=\sum_{m=0}^{q-1}u_m|\phi_{m} (X)\rangle$,
where $m\equiv(k_X+m,k_P)$ with $0\le k_X<1$ and $\phi_{m} (X)$ is the basis (\ref{Zakbasis}) of the $kq$-representation. In this subspace, the eigenequation is simply \cite{Brown1968}
\begin{equation}
2\cos\lambda(k_X+m)u_m+e^{-i\lambda k_P}u_{m-1}+e^{i\lambda k_P}u_{m+1}=\frac{4E}{K}u_m
\end{equation}
together with the boundary condition $u_0=u_q$.  To eliminate the dependence on $k_P$, we make the substitution $\bar{u}_m=e^{im\lambda k_P}u_m$,
which leads to
\begin{equation}
2\cos\lambda(k_X+m)\bar{u}_m+\bar{u}_{m-1}+\bar{u}_{m+1}=\frac{4E}{K}\bar{u}_m
\end{equation}
and the boundary condition $\bar{u}_0=e^{-i2\pi p k_P}\bar{u}_q$.
This equation can be formulated as
\begin{equation}
U_{m-1}=\mathbf{T}_mU_m
\end{equation}
where $U_m=(\bar{u}_m,\bar{u}_{m+1})$ and the matrix $\mathbf{T}_m$ is
\begin{equation}
\mathbf{T}_m=
\begin{bmatrix}
    \frac{4E}{K}-2\cos\lambda(k_X+m) & -1 \\
    1 & 0
\end{bmatrix}.
\end{equation}
From this recursive relation it is easy to find that $U_1=\mathbf{T}(k_X)U_q$ with $\mathbf{T}(k_X)=\prod\limits_{m=1}^q\mathbf{T}_m$. As $U_1=e^{-i2\pi p k_P}U_q,$ we have
the secular equation $\mathrm{det}(\mathbf{T}(k_X)-e^{-i2\pi p k_P})=0$, which can be expanded as
\begin{equation}\label{eq:F5}
\mathrm{Tr}\,\mathbf{T}(k_X)=2\cos(iq\lambda k_P).
\end{equation}
Due to the cyclic permutation invariance of matrix trace, we have $\mathrm{Tr}\,\mathbf{T}(k_X+1)=\mathrm{Tr}\,\mathbf{T}(k_X)$. Thus the expansion of
$\mathrm{Tr}\,\mathbf{T}(k_X)$ in terms of power series of $e^{i\lambda k_X}$ is simplely
\begin{equation}
\mathrm{Tr}\,\mathbf{T}(k_X)=C_0+C_q e^{iq\lambda k_X}+C_q^*e^{-iq\lambda k_X}.
\end{equation}
The coefficient $C_q$ can be directly evaluated by extracting the term with the highest power, resulting in $C_q=-1$. This leads to the relation
$$\mathrm{Tr}\,\mathbf{T}(k_X)=\mathrm{Tr}\,\mathbf{T}(0)+2-2\cos\lambda qk_X,$$ which gives Eq.~(\ref{eq:butterfly}) combining with Eq.~(\ref{eq:F5}).

\section{Kicking Dynamics}\label{app:kickingdynamics}

As we mentioned, the kicking dynamics is realized by an unitary transformation $\hat{\rho}_n^-\rightarrow\hat{\rho}_n^+=\hat{U}_K\hat{\rho}_n^-\hat{U}_K^\dagger$. To translate it in terms of
the characteristic function $w(s,k)$, we do the straightforward calculation
\begin{eqnarray}
&&w(s,k;\tau_n^+)\nonumber\\
&=&\int\,dx e^{ixk/\lambda}\langle x+s/2|\hat{U}_K\hat{\rho}_n^-\hat{U}_K^\dagger|x-s/2\rangle \nn
&=&\sum_{j=-\infty}^{\infty}J_j(2K\tau\sin\frac s2/\lambda)\int dxe^{ix(k+j\lambda)}\langle x+s/2|\hat{\rho}_n^-|x-s/2\rangle \nn
&=&\sum_{j=-\infty}^{\infty}J_j(2K\tau\sin\frac s2/\lambda)w(s,k+j\lambda;\tau_n^-),
\end{eqnarray}
where we have used the Jacobi-Anger relation in the second line.

\section{Obtaining $Q(X,P)$ from $w(s,k)$}\label{app:husimiQ}

The definition of the characteristic function $w(s,k)$ we used in the main text can be formulated into an elegant form \cite{Knight2005}
\begin{eqnarray}
w(\eta,\eta^*)=\mathrm{Tr}\,\rho e^{\eta a^\dagger-\eta^*a}
\end{eqnarray}
with $\eta=(s+ik)/\sqrt{2\lambda}$, whereas the characteristic function of the Husimi distribution is given by,
\begin{equation}
C_Q(\eta,\eta^*)=\mathrm{Tr}\,\rho e^{-\eta^*a}e^{\eta a^\dagger}.
\end{equation}
Clearly, the two characteristic functions are related through $C_Q(\eta,\eta^*)=w(\eta,\eta^*)e^{-|\eta|^2/2}$.
Once $C_Q(\eta,\eta^*)$ is obtained, the Husimi distribution $Q(X,P)$ can be retrieved by the Fourier transform
$$Q(\alpha,\alpha^*)=\frac{1}{\pi^2}\int d\eta^2e^{\alpha\eta^*-\alpha^*\eta}C_Q(\eta,\eta^*)$$ with $\alpha=(X+iP)/\sqrt{2\lambda}$.

\section{Expressions for $U_c$ and $U_e$}\label{app:UcUe}

First, we calculate the matrix element of $U(\hat{R})$ for two
given functions $f(X_1,X_2)$ and $g(X_1,X_2)$ as following
\begin{eqnarray}\label{URInner}
&&\langle f(X_1,X_2)|U(\hat{R})|g(X_1,X_2)\rangle\nonumber\\
&=&\langle f(X_c+\frac{1}{2}\Delta X,X_c-\frac{1}{2}\Delta
X)|U(\hat{R})|g(X_c+\frac{1}{2}\Delta X,X_c-\frac{1}{2}\Delta
X)\rangle\nonumber\\
&=&\sum_N U(R_N)\int dC \langle f(X_c+\frac{1}{2}\Delta
X,X_c-\frac{1}{2}\Delta X)|N,C\rangle\nonumber\\
&&\times\langle N,C|g(X_c+\frac{1}{2}\Delta
X,X_c-\frac{1}{2}\Delta
X)\rangle\nonumber\\
&=&\sum_NU(R_N)\int  dC \langle f(C+\frac{1}{2}\Delta
X,C-\frac{1}{2}\Delta X)|\Phi_N(\Delta X)\rangle\nonumber\\
&&\times\langle \Phi_N(\Delta X)|g(C+\frac{1}{2}\Delta
X,C-\frac{1}{2}\Delta X)\rangle.
\end{eqnarray}
Here, we have used the property of $\hat{X}_c|C\rangle=C|C\rangle$
and the resulting $\langle f(X_c)|C\rangle=\int
f(X_c)\delta(X_c-C) dX_c=f(C)$ for any $f(X_c)$ in the
representation of coordinate of center of mass. Then, we apply the
result (\ref{URInner}) to calculate $U_c$ and
$U_e$ defined in the main text, i.e., the direct integral
\begin{eqnarray}\label{DA}
U_c&=&\langle \varphi(X_1)\phi(X_2)|U(\hat{R})|\varphi(X_1)\phi(X_2)\rangle\nonumber\\
&=&\sum_NU(R_N)\int  dC \langle \varphi(C+\frac{1}{2}\Delta
X)\phi(C-\frac{1}{2}\Delta X)|\Phi_N(\Delta X)\rangle\nonumber\\
&&\times\langle \Phi_N(\Delta X)|\varphi(C+\frac{1}{2}\Delta
X)\phi(C-\frac{1}{2}\Delta X)\rangle\nonumber\\
&=&\sum_NU(R_N)\int  dC \Big|\langle \varphi(C+\frac{1}{2}\Delta
X)\phi(C-\frac{1}{2}\Delta X)|\Phi_N(\Delta
X)\rangle\Big|^2\nonumber\\
&\equiv&\sum_NU(R_N) I_N,
\end{eqnarray}
and the exchange integral
\begin{eqnarray}\label{QA}
U_e&=&\langle \varphi(X_1)\phi(X_2)|U(\hat{R})|\phi(X_1)\varphi(X_2)\rangle\nonumber\\
&=&\sum_NU(R_N)\int  dC \langle \varphi(C+\frac{1}{2}\Delta
X)\phi(C-\frac{1}{2}\Delta X)|\Phi_N(\Delta X)\rangle\nonumber\\
&&\times\langle \Phi_N(\Delta X)|\phi(C+\frac{1}{2}\Delta
X)\varphi(C-\frac{1}{2}\Delta X)\rangle\nonumber\\
&=&\sum_N(-1)^NU(R_N)\nonumber\\
&&\times\int  dC \Big|\langle \varphi(C+\frac{1}{2}\Delta
X)\phi(C-\frac{1}{2}\Delta
X)|\Phi_N(\Delta X)\rangle\Big|^2\nonumber\\
&\equiv&\sum_N(-1)^NU(R_N) I_N.
\end{eqnarray}
Here, the overlap integral $I_N$ is given by
\begin{eqnarray}
I_N\equiv \int  dC \Big|\langle \varphi(C+\frac{1}{2}\Delta
X)\phi(C-\frac{1}{2}\Delta X)|\Phi_N(\Delta X)\rangle\Big|^2.
\end{eqnarray}

\section{$I_N$ for coherent states}\label{app:CSR}

Now, we assume the two states $\varphi(X)$ and $\phi(X)$
are two displaced squeezed coherent states described by
\begin{eqnarray}\label{2localstates}
\left\{
\begin{array}{lll}
\varphi(X)&=&\psi_0(X-r_m)=\Big(\frac{\beta}{\sqrt{\pi}}\Big)^{\frac{1}{2}}e^{-\frac{1}{2}\beta^2(X-r_m)^2}\\
\phi(X)&=&\psi_0(X+r_m)=\Big(\frac{\beta}{\sqrt{\pi}}\Big)^{\frac{1}{2}}e^{-\frac{1}{2}\beta^2(X+r_m)^2}.\\
\end{array}
\right.
\end{eqnarray}
The product state
$\varphi(X_1)\phi(X_2)=\varphi(C+\frac{1}{2}\Delta
X)\phi(C-\frac{1}{2}\Delta X)$ can be calculated from Eq.~(\ref{2localstates})
\begin{eqnarray}\label{}
\varphi(X_1)\phi(X_2)
&=&\frac{\beta}{\sqrt{\pi}}e^{-\beta^2C^2}e^{-\beta^2(\frac{1}{2}\Delta
X-r_m)^2}.
\end{eqnarray}
From Eq.~(\ref{overlapIN}), we obtain the following overlap integral
\begin{eqnarray}\label{INa}
I_N &&=\frac{\beta^2}{\pi}\int_{-\infty}^{+\infty}
e^{-2\beta^2C^2}dC \Big|\langle e^{-\beta^2(\frac{1}{2}\Delta
X-r_m)^2}|\Phi_N(\Delta
X)\rangle\Big|^2\nonumber\\
&&=\Big|\langle\Phi_N(\Delta
X)|\Big(\frac{\beta}{\sqrt{2\pi}}\Big)^{\frac{1}{2}}
e^{-\frac{1}{2}(\frac{\beta}{\sqrt{2}})^2(\Delta
X-2r_m)^2}\rangle\Big|^2.
\end{eqnarray}
The displacement operator $\hat{D}_\gamma=\exp\Big(\gamma \hat{a}^\dagger-\gamma^*
\hat{a}\Big)$ has the property $\hat{D}_\gamma^\dagger
\hat{a} \hat{D}_\gamma=\hat{a}+\gamma$. We further introduce the
squeezing operator $$\hat{S}_\xi\equiv\exp\Big[\frac{1}{2}(\xi^*
\hat{a}^2-\xi \hat{a}^{\dagger 2})\Big]$$ with parameter $\xi=re^{i\theta}$.
The squeezing operator has the following property $$\hat{S}_\xi^\dagger \hat{a}
\hat{S}_\xi=v\hat{a}+u\hat{a}^\dagger$$ with the squeezing parameters $v=\cosh r,\ u=-e^{i\theta} \sinh r$.
Inversely, the parameter $\xi=re^{i\theta}$ is related to $v$ and $u$ via
\begin{eqnarray}\label{}
\left\{
\begin{array}{lll}
r&=&\mathrm{arccosh}(v)=\ln(v+\sqrt{v^2-1})\\
e^{i\theta}&=&-u/\sinh r.\\
\end{array}
\right.
\end{eqnarray}
Using operators $\hat{D}_\gamma$ and $\hat{S}_\xi$, we write the displaced
squeezed state in Eq.(\ref{INa}) as
\begin{eqnarray}\label{}
\Big|\Big(\frac{\beta}{\sqrt{2\pi}}\Big)^{\frac{1}{2}}
e^{-\frac{1}{2}(\frac{\beta}{\sqrt{2}})^2(\Delta
X-2r_m)^2}\Big\rangle=\hat{D}_{-\gamma}\hat{S}_{-\xi}|0\rangle\equiv|-\gamma,-\xi\rangle,\ \
\end{eqnarray}
where the displacement parameter is $\gamma=-{r_m}/{\sqrt{\lambda}}$
and the squeezing parameters are given by
\begin{eqnarray}\label{parametes}
v=\frac{1}{2}\Big(\sqrt{\lambda}\ \beta+\frac{1}{\sqrt{\lambda}\
\beta}\Big), \ \ \  u=\frac{1}{2}\Big(\sqrt{\lambda}\
\beta-\frac{1}{\sqrt{\lambda}\ \beta}\Big).
\end{eqnarray}
Using the formula (7.81) in Ref.~\cite{Knight2005}, the overlap integral $I_N=|\langle N|-\gamma,-\xi\rangle|^2$ is given by
\begin{eqnarray}\label{INb}
I_N &=&\frac{(\frac{1}{2}\tanh r)^N}{N!\cosh
r}\Big|H_N\Big[\frac{\gamma^*e^{i\theta}\sinh
r-\gamma\cosh r}{\sqrt{e^{i(\theta+\pi)}\sinh
(2r)}}\Big]\Big|^2\nonumber\\
&&\times\exp\Big[-|\gamma|^2+\frac{1}{2}(\gamma^{*2}e^{i\theta}+
\gamma^2e^{-i\theta})\tanh r\Big]. \ \ \
\end{eqnarray}
Given the parameters $r_m$ and $\beta$,
we can calculate $U_c$ and $U_e$ from
Eq.s~(\ref{DQ}), (\ref{parametes}) and (\ref{INb}).

The standard coherent state, whose squeezing parameters are $v=1$ and $u=0$, can be obtained
by choosing $\beta=1/\sqrt{\lambda}$ in Eq.~(\ref{parametes}). From Eq.(\ref{INb}), the overlap
integral of two standard coherent states $I^s_N$ can be calculated
\begin{eqnarray}\label{INSC}
I^s_N
=\frac{|\gamma|^{2N}}{N!}e^{-|\gamma|^2}=\frac{1}{N!}\Big(\frac{R^2}{4\lambda}\Big)^N\exp\Big(-\frac{R^2}{4\lambda}\Big).
\end{eqnarray}
Here, $
R\equiv 2r_m=2\gamma\sqrt{\lambda}
$ is the distance between the centers of two
coherent states in phase space.
The quantity $R$ is different from the quantized phase space
distance $R_N$. For two overlapped coherent states, their distance
$R$ is zero but $R_N$ is always positive as shown by
Eq.~(\ref{QuantizedRN}).

\section{ $U(R_N)$ for hardcore interaction }\label{app:hardcore}
We derive $U(R_N)$ for hard-core interaction in Eq.~(\ref{UNhardcore}). In the rest frame, assuming the interaction potential between two atoms is $V(x_1-x_2)$, the eigen equation of energy is given by
\begin{eqnarray}\label{ET}
&&\Big[-\frac{\lambda^2}{2}\frac{\partial^2}{\partial x_1^2}-\frac{\lambda^2}{2}\frac{\partial^2}{\partial x_2^2}
+\frac{1}{2}(x^2_1+x^2_2)+V(x_1-x_2)\Big]\Psi(x_1,x_2)\nonumber\\
&&=E_T\Psi(x_1,x_2).
\end{eqnarray}
Herr, $E_T$ is the total energy. We introduce the coordinate of
central of mass $x_c\equiv(x_1+x_2)/2$ and the relative coordinate
$x\equiv x_1-x_2$. By separating the eigenstate into a product state $\Psi=\phi(x_c)\psi(x)$, we have the eigen
equation for the motion of center of mass
\begin{eqnarray}\label{eigencenter}
\Big(-\frac{\lambda^2}{2M}\frac{\partial^2}{\partial x_c^2}+\frac{1}{2}Mx_c^2\Big)\phi(x_c)&=&E_c\phi(x_c)
\end{eqnarray}
with $M=2$ the total mass and $E_c$ the energy
of center of mass motion. The eigen equation for the relative motion is
\begin{eqnarray}\label{eigenrelative}
\Big[-\frac{\lambda^2}{2\mu}\frac{\partial^2}{\partial x^2}+\frac{1}{2}\mu x^2+V(x)\Big]\psi(x)&=&E\psi(x)
\end{eqnarray}
with $\mu=1/2$ the reduced mass, $E=E_T-E_c$ is the energy of relative motion.

The solutions of Eq.~(\ref{eigencenter}) are just the harmonic motions. We now try to find the solutions of Eq.~(\ref{eigenrelative}). Without interaction $V(x_1-x_2)$, the eigen problem is determined by  $H_r\phi_n(x)=E_n\phi_n(x)$ with $$H_r\equiv -\frac{\lambda^2}{2\mu}\frac{\partial^2}{\partial x^2}+\frac{1}{2}\mu x^2.$$ The eigenstates are given by
\begin{eqnarray}\label{}
\psi_n(x) =\Big(\frac{\zeta}{\sqrt{\pi} 2^n
n!}\Big)^{\frac{1}{2}}H_n\Big(\zeta x\Big)e^{-\frac{1}{2}\zeta^2x^2},
\end{eqnarray}
where the parameter $\zeta=\sqrt{1/(2\lambda)}$. With consideration of hard-core interaction, i.e., $V(x_1-x_2)=+\infty$ for $|x_1-x_2|<2a$ and $V(x_1-x_2)=0$ for $|x_1-x_2|>2a$, the boundary condition requires that wavefunction must be zero at $x\in [-a,a]$. For odd integer $n$, we assume the approximate eigenstates are just repulsed outside the hard-core region, i.e.,
\bea
\phi_{2n+1}(x)=
\begin{cases}
\psi_{2n+1}(x-2a)  &  x\ge 2a    \\
0   &  -2a<x<2a  \\
\psi_{2n+1}(x+2a)  &  x<-2a
\end{cases},
\qquad n\in\mathbb{N} \qquad
\eea
For even integer $n$, the wave functions $\phi_n(x)$, however, do not satisfy the hard-core boundary condition and the continuity condition. Therefore, we construct the symmetric eingenstates from antisymmetric states
\bea
\phi_{2n}(x)=
\begin{cases}
\psi_{2n+1}(x-2a)  &  x\ge 2a    \\
0  &  -2a<x<2a  \\
-\psi_{2n+1}(x+2a)  &  x<-2a
\end{cases}.
\qquad n\in\mathbb{N} \qquad
\eea

The energy levels to the first order correction are
\begin{widetext}
\bea\label{1stordercorrection}
&&\langle\phi_{2n}|H_r|\phi_{2n}\rangle=\langle\phi_{2n+1}|H_r|\phi_{2n+1}\rangle   \nonumber    \\
&&=\frac12\int_{-\infty}^{-2a}dx\,\psi_{2n+1}^*(x+2a)\Big(-\frac{\lambda^2}{\mu}\frac{\partial^2}{\partial x^2}+\mu x^2\Big)\psi_{2n+1}(x+2a)+\frac12\int_{2a}^{\infty}dx\,\psi_{2n+1}^*(x-2a)\Big(-\frac{\lambda^2}{\mu}\frac{\partial^2}{\partial x^2}+\mu x^2\Big)\psi_{2n+1}(x-2a)   \nonumber        \\
&&=\int_{-\infty}^{\infty}dx\,\psi_{2n+1}^*(x)H_r\psi_{2n+1}(x)+2\mu^2a^2\int_{-\infty}^{\infty}dx\,\psi_{2n+1}^*(x)\psi_{2n+1}(x)+4\mu a\int_0^\infty dx\,\psi_{2n+1}^*(x)\,x\,\psi_{2n+1}(x)   \nonumber    \\
&&=\lambda(2n+1)+\frac{1}{2}a^2   + \frac{a}{\zeta}\frac{1}{(2n+1)!2^{2n+1}\sqrt{\pi}}\sum_{k,l=0}^{[n+\frac12],[n+\frac12]}\frac{(-1)^{k+l}[(2n+1)!]^2(2n+1-k-l)!}{k!l!(2n+1-2k)!(2n+1-2l)!}2^{2(2n+1-k-l)},\ \ \ \ \ \ \
\eea
\end{widetext}
where we have used $\mu=1/2$ in the last step. In fact, one can prove that
$U(R_N)$ is the first order correction, from the weak interaction, to the $N$-th energy level of the harmonic trapping potential. Relabelling $N\equiv2n+1$, we have from Eq.~(\ref{1stordercorrection})
\bea\label{}
U(R_N)=\frac{a\sqrt{2\lambda/\pi}}{2^{-N} N!}\sum_{k,l=0}^{[\frac N2][\frac N2]}\,\frac{(-1)^{k+l}(N!)^2(N-k-l)!}{2^{2k+2l}k!l!(N-2k)!(N-2l)!},\ \ \ \ \
\eea
which is exactly Eq.~(\ref{UNhardcore}) in the Sec.~\ref{IVB}.


\begin{thebibliography}{123}
\bibitem [{sho()}]{short-range}%
  \BibitemOpen
  \href@noop {} {\ }\bibinfo {note} {In the rigorous proof of Mermin-Wagner
  theorem, it is enough that $\sum_{\vec{R}}R^2|U_e(\vec{R})|$
  converge.}\BibitemShut {Stop}%
\end{thebibliography}
%

\end{document}